\documentclass[preprint]{ptephy_v1}

\preprintnumber{XXXX-XXXX} 

\usepackage{multirow}
\usepackage{graphicx}
\usepackage{afterpage}
\usepackage[nodayofweek]{datetime}
\settimeformat{hhmmsstime}
\usepackage{textcomp}
\usepackage{threeparttable}
\usepackage{arydshln}
\usepackage{tablefootnote}
\usepackage[capitalize]{cleveref}
\usepackage{comment}
\usepackage{lineno}
\usepackage{here}

\Crefname{figure}{Fig.}{Figs.}

\title{Study of muonium emission from laser-ablated silica aerogel}

\author{J.~Beare}
\affil{Department of Physics and Astronomy, McMaster University, ON L8S 4L8, Canada}

\author{G.~Beer}
\affil{Department of Physics and Astronomy, University of Victoria, BC V8P 5C2, Canada}

\author{J.H.~Brewer}
\affil{Department of Physics and Astronomy, University of British Columbia, BC V6T 1Z1, Canada}

\author[4,5]{T.~Iijima}
\affil{Kobayashi-Maskawa Institute, Nagoya University, Aichi 464-8602, Japan}
\affil[5]{Graduate School of Science, Nagoya University, Aichi 464-8602, Japan}

\author[6]{K.~Ishida}
\author[6]{M.~Iwasaki}
\affil[6]{RIKEN Nishina Center, RIKEN, Saitama 351-0198, Japan}

\author[7]{S.~Kamal}
\affil[7]{Laboratory for Advanced Spectroscopy and Imaging Research (LASIR), Department of Chemistry, University of British Columbia, BC V6T 1Z1, Canada}

\author[8]{K.~Kanamori}
\affil[8]{Department of Chemistry, Graduate School of Science, Kyoto University, Kyoto 606-8502, Japan}

\author[9]{N.~Kawamura}
\affil[9]{High Energy Accelerator Research Organization (KEK), Ibaraki 305-0801, Japan}

\author[10]{R.~Kitamura}
\affil[10]{Japan Atomic Energy Agency (JAEA), Ibaraki 319-1195, Japan}

\author[11]{S.~Li}
\affil[11]{Department of Physics, University of Tokyo, Tokyo 113-8654, Japan}

\author[1]{G.M.~Luke}

\author[12]{G.M.~Marshall}
\affil[12]{TRIUMF, BC V6T 2A3, Canada}

\author[9]{T.~Mibe}
\author[9]{Y.~Miyake}
\author[9]{Y.~Oishi}
\author[12]{K.~Olchanski}
\author[2,12]{A.~Olin}
\author[9]{M.~Otani}

\author[13]{M.A.~Rehman}
\affil[13]{The Graduate University for Advanced Studies, Kanagawa 240-0193, Japan}

\author[9,11,14]{N.~Saito}
\affil[14]{J-PARC Center, Ibaraki 319-1195, Japan}

\author[9]{Y.~Sato}
\author[9]{K.~Shimomura}
\author[9,*]{K.~Suzuki}

\author[15]{M.~Tabata}
\affil[15]{Department of Physics, Chiba University, Chiba 263-8522, Japan
\email{kazuhito.suzuki@j-parc.jp}}

\author[11]{H.~Yasuda\thanks{These authors contributed equally to this work}}

\subjectindex{C31, G04}

\begin{document}

\begin{abstract}
The emission of muonium ($\mu^+e^-$) atoms into vacuum from silica aerogel with laser ablation on its surface was studied with various ablation structures at room temperature using the subsurface muon beams at TRIUMF and Japan Proton Accelerator Research Complex (J-PARC). Laser ablation was applied to produce holes or grooves with typical dimensions of a few hundred $\mu$m to a few mm, except for some extreme conditions. The measured emission rate tends to be higher for larger fractions of ablation opening and for shallower depths. More than a few ablation structures reach the emission rates similar to the highest achieved in the past measurements. The emission rate is found to be stable at least for a couple of days. Measurements of spin precession amplitudes for the produced muonium atoms and remaining muons in a magnetic field determine a muonium formation fraction of $(65.5 \pm 1.8)$\%.
The precession of the polarized muonium atoms is also observed clearly in vacuum.
A projection of the emission rates measured at TRIUMF to the corresponding rates at J-PARC is demonstrated taking the different beam condition into account reasonably.

\end{abstract}

\maketitle

\section{Introduction}

Applications of intense beams of muons with
energies below a few tens of keV (``keV muons'') are
growing in a variety of scientific fields.
Low energy muons can be excellent probes to investigate the
magnetic properties of materials using muon spin relaxation
($\mu$SR) techniques~\cite{muSR}. Applications can be extended to thin layers
at a scale less than 10 nm for energies below 1 keV.
Low energy muons can also be utilized in the form of muonium
($\mu^+e^-$, or Mu) atoms, both as an analogue to a light isotope of
hydrogen for surface chemistry~\cite{SurfaceChem} and as a purely leptonic bound state for precision tests of quantum electrodynamics (QED)~\cite{MuLambShift,MuHFS,Mu1S2S}.
Furthermore, an intense muon beam with extremely low energy
naturally has a small phase space volume, thus it can provide a
viable small emittance source for subsequent acceleration.  The
resulting muon beam of high intensity retains a small emittance,
offering the possibility to extend these applications as well as
open up new windows into particle physics with potential for
revolutionary precision measurements.

Positive muon beams of a few MeV energy are typically obtained
from the decay of positive pions stopped at or near the surface
of a pion production target. These are referred to as surface
muons; negative pions are captured by the target nuclei promptly
and do not contribute.  At their point of production, surface
muons have a monochromatic energy (momentum) of 4.1 MeV (29.8
MeV/$c$) with 100\% spin-polarization anti-parallel to the
momentum direction due to the parity-violating
$\pi^+\to\mu^+\nu_{\mu}$ decay. The conversion of MeV muons to a
small emittance source of keV muons requires a cooling process.
Because of the short muon lifetime (2.2 $\mu$s), muon cooling is
challenging and various innovative methods are under
development~\cite{LEM, MuCool, MICE}.
One such method is to thermalize the surface muons in the form of
muonium atoms in an appropriate material, then allow them to
diffuse into vacuum where they can be ionized with a laser
resonant ionization technique~\cite{LaserRes,LaserResRAL}.  The
resulting positive muons are effectively cooled to an energy
distribution characterized by the material temperature.
Typically 50\% of the muon beam's polarization is lost due to the
hyperfine interaction with initially unpolarized electrons of the
material during muonium formation. The material can ideally be
chosen for high muonium formation probability, low additional
depolarization, and high efficiency of muonium diffusion.
A tungsten foil heated to 2000 K was used in Refs.~\cite{LaserRes,LaserResRAL}, from which muonium was emitted at an energy of about 0.2 eV.

Development of this muon cooling method using a more advanced ionization laser system~\cite{LymanAlpha} has continued at a muon beamline~\cite{Uline} of Materials and Life Science Experimental Facility (MLF) at Japan Proton Accelerator Research Complex (J-PARC).
It is proposed to extend the method further in the J-PARC muon $g$-2/EDM experiment~\cite{E34}, which is now under construction. The goal is to measure the muon anomalous magnetic moment and electric dipole moment with a totally new approach based on a small-emittance muon beam.
Here, silica aerogel has been chosen for the material of the muonium production target since it can produce thermalized polarized muonium atoms, has a porous structure that allows diffusion, and is compatible with the accelerator's vacuum requirements. 
The surface muon beam is to be thermalized at room temperature to the energy (momentum) of about 25 meV (2.3 keV/$c$). The resulting thermal muonium atoms in vacuum are ionized and the muons resulting from ionization are re-accelerated up to 212 MeV (300 MeV/$c$), 
achieving a total momentum spread of 0.04\% (RMS) with a transverse emittance of 0.33$\pi$ mm mrad~\cite{E34} which is unprecedentedly small for a muon beam.
Past studies demonstrated muonium production in silica aerogel
and its emission into vacuum~\cite{PTEP2013} and observed a
significant improvement of the emission rate with laser ablation
on the aerogel surface~\cite{PTEP2014}.
The present issues are to understand better the underlying processes leading to muonium
emission with laser ablation, to find the optimal ablation
structure, and to confirm the polarization of muonium following
emission into vacuum.

To address these issues, various ablation structures were examined
with the surface muon beams at TRIUMF and J-PARC in 2017.
The TRIUMF muon beam was continuous with a momentum spread of about 2\% (RMS) while the J-PARC beam was pulsed with a spread of about 4\% (RMS).
Most of the ablation structures were examined at TRIUMF utilizing the smaller beam momentum spread, in the presence of a magnetic field where the effects of polarization would be apparent.
Some of them were also measured at J-PARC to demonstrate the projection of the measurements at TRIUMF into the expectations at J-PARC for the application to the muon $g$-2/EDM experiment.
This report describes the results of these measurements and projections on the muonium emission from laser-ablated silica aerogel targets.

\section{Muonium production targets}
\subsection{Target properties}

The muonium production targets are prepared in the form of slabs  50 mm $\times$ 50 mm in area and typically about 9 mm in thickness. Surface muons collimated to a beam diameter of 10 mm irradiate the large face of the target, which is suspended in a vacuum chamber.
The laser ablation is applied on the central 30 mm $\times$ 30 mm surface of the downstream large face, from which some of the produced muonium atoms and penetrating muons exit the target.

Most of the targets are made of silica aerogel~\cite{SilicaAerogel} with a density of $23.2~{\rm mg/cm}^3 - 23.6$ mg/cm$^3$ and a thickness of $8.7~{\rm mm} - 8.9$ mm.
The silica aerogel was rendered hydrophobic by surface modification with trimethylsilyl groups (Si(CH$_3$)$_3$) during the production process.
A different type of material, referred to as poly(methylsilsesquioxane)
(PMSQ) gel~\cite{PMSQ}, is also used to explore possible variations of the target material.
It is a novel organic-inorganic hybrid aerogel material, in which the silicon dioxide groups (SiO$_2$) are partially replaced with the methyl groups (CH$_3$). Its mechanical properties might be advantageous for the long-term mechanical stability compared to the fragile silica aerogel.
The PMSQ gels were produced in two different drying processes, the super-critical drying process  producing lower-density samples (52.5 mg/cm$^3$) and  the evaporation process under the ambient condition  producing higher-density samples (99.6 mg/cm$^3$).
These samples are produced with thicknesses of 4 mm and 2 mm, respectively, to have a similar target mass with the silica aerogel targets.
A target with no ablation, referred to as a ``flat'' target, is also prepared for each of the three materials as reference.
A silica plate with the thickness of 0.1 mm is used for calibration and background evaluation utilizing its high density and small thickness as described in later sections.

\begin{threeparttable}[htbp]
\caption{
Summary of target thickness ($T$), density ($\rho$), and the ablation structure parameters, hole diameter ($\phi$), groove width ($w$), pitch ($p$) and depth ($d$), used in the measurements at TRIUMF and J-PARC. The uncertainty of each parameter is uniform over the targets and described in the text.
The corresponding information for the targets used in the previous study~\cite{PTEP2014} is also listed for comparison.
}
\label{tab_targets}
\centering
\small
\begin{tabular}{cccccccl}
\hline\hline
Ablation & Target & $T$ [mm] & $\rho$ [mg/cm$^3$] & $\phi$ or $w$ [mm] & $p$ [mm] & $d$ [mm] & \multicolumn{1}{c}{Remark} \\
\hline\hline
Flat  & SP\tnote{a,c}  & 0.1 & $2.2\times 10^3$ & 0 & 0 & 0 & Silica plate \\ 
\hline
Flat & S08\tnote{a} & 8.8 & 23.5 & 0 & 0 & 0 \\
\cdashline{1-1}
\multirow{15}{*}{Holes}
& S01\tnote{a} & 8.8 & 23.6 & 0.25 & 0.4 & 1 \\
& S03\tnote{b} & 8.8 & 23.5 & 0.25 & 0.4 & 3 \\ 
& S04\tnote{a} & 8.8 & 23.6 & 0.25 & 0.4 & 5 \\
& S05\tnote{a} & 8.8 & 23.6 & 0.25 & 0.4 & 8.8 & Through holes \\
& S09 & 8.9 & 23.3 & 0.1 & 0.15 & 1.25 \\
& S10 & 8.9 & 23.3 & 0.1 & 0.2 & 1.5 \\
& S11 & 8.9 & 23.6 & 0.1 & 0.3 & 1.75 \\
& S12 & 8.9 & 23.6 & 0.1 & 0.5 & 1.75 \\
& S13 & 8.9 & 23.6 & 0.1 & 0.3 & 4 \\
& S18 & 8.8 & 23.2 & 0.165 & 0.25 & 1 \\
& S19 & 8.8 & 23.2 & 0.25 & 0.3 & 1.5\tnote{d} \\
& S20 & 8.8 & 23.2 & 0.07 & 0.105 & 1 \\
& S2013-1\tnote{c} & 8 & 26.0 & 0.27 & 0.4 & 4.75 & Reuse of O2013-3 \\
& S23 & 8.7 & 23.3 & 0.5 & 0.8 & 1.5 & Enhanced cones \\
\cdashline{1-1}
\multirow{3}{*}{Grooves}
& S21 & 8.8 & 23.2 & 0.25 & 0.35 & 1 & Channels \\
& S22 & 8.7 & 23.3 & 0.5 & 0.62 & 0.6\tnote{d} & Channels\\
& S24 & 8.7 & 23.3 & 0.22 & 0.3 & 1.5 & V-grooves \\
\cdashline{1-1}
Continuous & S14 & 8.9 & 23.6 & 30 & 30 & 0.4 \\
\hline
Flat  & S15 & 4 & 52.5 & 0    & 0    & 0
& \multirow{2}{*}{PMSQ gel (low-$\rho$)} \\
Holes & S16 & 4 & 52.5 & 0.25 & 0.4  & 0.5 \\
\cdashline{8-8}
Flat  & S17 & 2 & 99.6 & 0    & 0    & 0
& \multirow{2}{*}{PMSQ gel (high-$\rho$)} \\
Holes & S25 & 2 & 99.6 & 0.13 & 0.28 & 0.3 \\
\hline
Flat  & O2013-1 & 7 & 29 & 0    & 0   & 0
& \multirow{4}{*}{Ref.~\cite{PTEP2014}} \\
\cdashline{1-1}
\multirow{3}{*}{Holes}
& O2013-2 & 7 & 29 & 0.27 & 0.5 & 4.75 \\
& O2013-3 & 7 & 29 & 0.27 & 0.4 & 4.75 \\
& O2013-4 & 7 & 29 & 0.27 & 0.3 & 4.75 \\
\hline\hline
\end{tabular}
\begin{tablenotes}
\item[a]{Used in both the TRIUMF and J-PARC measurements.}
\item[b]{Used in the J-PARC measurements only.}
\item[c]{These targets have a height and width of 40 mm and 30 mm, respectively.}
\item[d]{The ablation depth varies $\pm 0.5$ mm for S19 and $\pm 0.1$ mm for S22.}
\end{tablenotes}
\end{threeparttable}

\vspace*{\baselineskip}

Laser ablation was introduced to produce a structure on the emission face that enhances muonium diffusion into the vacuum. This is based on the interpretation for the time evolution of the muonium population in vacuum obtained in the past study~\cite{PTEP2013}. The time evolution is reasonably described by a Monte Carlo simulation assuming muonium diffusion inside the porous silica aerogel with a Maxwell-Boltzmann three-dimensional velocity distribution. The simulation indicates that the muonium emission is limited by its diffusion length of about 30 $\mu$m; which is much smaller than the range straggling of stopping muons in a target, about 2 mm (FWHM). This interpretation suggests a possible increase of the emission rate by having a structure with dimensions of order 0.1 mm on the emission face. Here, the muon stopping distribution is assumed to be peaked at or near the emission face by tuning the beam momentum of the incident surface muons.

The muonium emission rate of an aerogel target with holes on the emission face with the diameter of 0.27 mm, pitch of 0.3 mm and depth ranging in $4.5~{\rm mm}-5.0$ mm was about 8 times higher than that of the flat target for the silica aerogel with the density of 29 mg/cm$^3$~\cite{PTEP2014}.
The same ablation structure with larger pitches also obtained higher emission rates than the flat target did, but the increase tends to decrease for larger pitches. A trend indicating more muonium emission with a denser structure was less clear.
Naive considerations can lead to the hypothesis that the muonium emission is dominated by the ablated surface area. In this case, the hole diameter and depth will also influence the emission rate and need to be surveyed, too.
The structure geometry can also influence the emission rate. A hole structure largely contains the ablated surface inside the holes while grooves have an open structure in one direction. Extending the open structure in two directions leads to the continuously ablated area.

The properties of the targets studied are summarized in Tables~\ref{tab_targets}.
All the targets were used in the TRIUMF measurements, except the target S03 which was used in the J-PARC measurement only. The other targets of S01$-$S08 were used in both measurements.
The target thickness is measured at an edge using a linear scale with the resolution of 0.25 mm while the  weight is measured using an electronic scale with the resolution of 2 mg. Thus the  target densities listed have  uncertainties of $\pm 0.7$ mg/cm$^3$.
The ablation structure parameters are the hole diameter ($\phi$) or groove width ($w$), pitch ($p$) and depth ($d$) as illustrated in \Cref{fig:ablation_aspect_ratio,fig:ablation_cross_section,fig:channel_pattern}.
The holes are arranged two-dimensionally in an equilateral-triangular configuration at the vertices with the side length corresponding to a given pitch. The grooves run over the area in one direction and are arranged in parallel with a given pitch.
The uncertainties of the structure parameters are considered to be $\pm 20$\% for the diameter and width, $\pm 5$ $\mu$m for the pitch and $\pm 0.25$ mm for the depth as described in Sec.~\ref{sec:laser}.

\subsection{Laser ablation technique}
\label{sec:laser}
The aerogel targets were ablated with ultra-short laser pulses from an amplified Ti:Sapphire femtosecond laser~\cite{Coherent} with the specification listed in Table~\ref{tab:laser_spec}.
We also examined several other pulsed laser sources for laser ablation, including nanosecond and picosecond lasers at 355 nm and 532 nm, and excimer laser at 193 nm. It was found that with femtosecond pulses, the ablated structures were much better defined and had significantly less peripheral damage. The use of femtosecond laser also allowed for rather fine control over the depth of ablated structures, from few hundreds $\mu$m to several mm.
Using a 300 mm focal length lens, the laser beam was focused on the surface of the target to a typical spot size of about 40 $\mu$m, with sufficient fluence to exceed the ablation threshold of about 2 mJ/cm$^2$.
Computer-controlled translation stages moved the target under the focused laser beam to produce a pre-programmed ablation structure, while an electronic shutter controlled the laser exposure time, or in effect, the number of laser pulses striking the surface.
The pulse energy was adjusted using neutral density filters, and the laser spot size was increased if needed by reducing the diameter of the incident laser beam at the lens through a partially closed aperture.
These controls were used to produce a series of structures: arrays of holes with various diameter, pitch, and depth; one-dimensional grooves; and an area of two-dimensional continuous ablation.
\begin{table}[!htbp]
\caption{
Laser specification used for the ablation. The scan speed was set only for groove and continuous ablation.
}
\label{tab:laser_spec}
\centering
\small
\begin{threeparttable}
\begin{tabular}{lcc}
\hline\hline
\multicolumn{1}{c}{Item} & Value & Unit\\
\hline\hline
Wavelength & 800 & nm\\
Pulse duration & 120 & fs\\
Pulse energy & $40-400$ & $\mu$J \\
Repetition rate & $0.1-1$ & kHz\\ 
Spot size at the focus & $6-110$ & $\mu$m \\
Exposure time & $30-150$ & ms\\
Scan speed & $1-2.5$ & mm/s\\
\hline\hline
\end{tabular}
\end{threeparttable}
\end{table}

\begin{figure}[!htbp] 
\begin{center} 
\includegraphics[width=0.8\columnwidth]{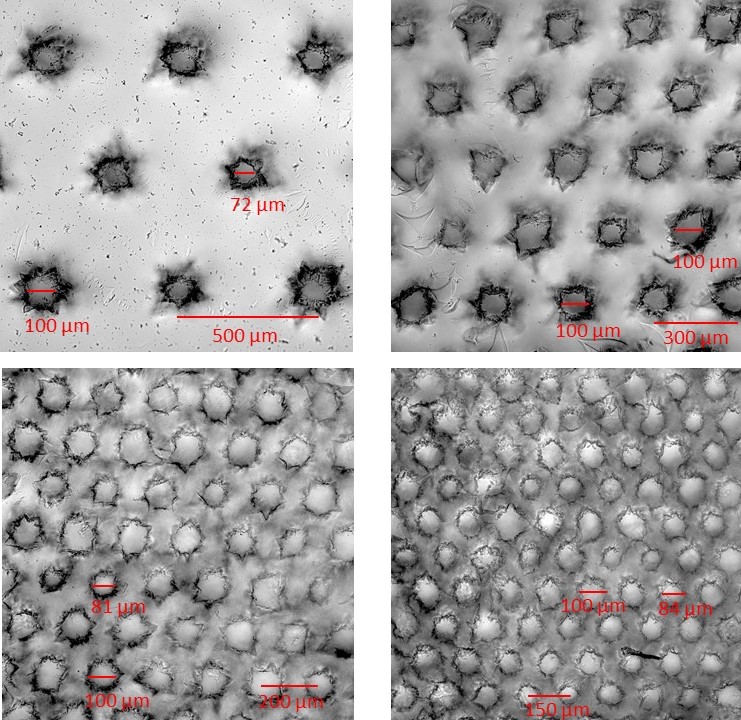} 
\caption{
Microscopic pictures of aerogel targets with hole ablation obtained using a laser-scanning confocal microscope as described in the text. All the hole structures have a diameter of about 100 $\mu$m. From the top left to the bottom right, the pitch is 500 $\mu$m, 300 $\mu$m, 200 $\mu$m and 150 $\mu$m. 
} 
\label{fig:ablation_aspect_ratio} 
\end{center} 
\end{figure}

A laser pulse induces a plasma in a material and springs off the local material, giving an opening with non-smooth envelope and irregular shape as well as micro-cracks in the surroundings. Multiple laser pulses at the same position grow the opening in depth, resulting in a hole.
Microscopic pictures of the examples of laser-ablated holes are shown in Fig.~\ref{fig:ablation_aspect_ratio}. These images were acquired by the use of a laser-scanning confocal microscope~\cite{Zeiss}, in which a 543 nm He-Ne laser scans across the surface of the target and the transmitted beam is recorded with a photo-multiplier tube to form the image. The image stack in the depth direction was utilized to obtain three-dimensional views of the hole structure.

\begin{figure}[htb] 
\begin{center} 
\includegraphics[width=0.8\columnwidth]{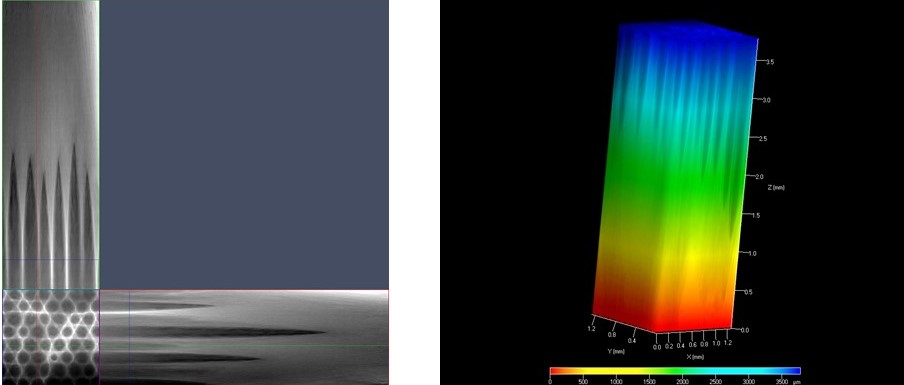} 
\caption{
Cross-sectional views of an ablated hole structure laid out with the front view (left), and their  three-dimensional contour image (right). The images are obtained using a laser-scanning confocal microscope as described in the text.
} 
\label{fig:ablation_cross_section} 
\end{center} 
\end{figure}

The pitch of the holes is controlled within $\pm 5$ $\mu$m using computer-controlled translation stages.
The hole diameter was controlled by tuning the laser pulse energy and exposure time, however, typically about $\pm$20\% variation in the hole diameter was observed.
The depth of the holes was primarily controlled by the number of laser pulses irradiated onto the surface. This was achieved by an electronic shutter. For shallow holes of 1 mm depth, as few as 10 laser pulses were needed per hole. The depth of the holes varied by about $\pm 250$ $\mu$m under the same ablation conditions.
Given the focal point is fixed at the target surface, the laser impact is less at a deeper position in the target, having more influence by the local sparseness and density of the target material and less control on the depth. It also results in a pointing shape at the bottom of the hole as shown in the cross-sectional view of Fig.~\ref{fig:ablation_cross_section}. Therefore, the hole shape is considered to be conical.
The hole structure with an enhanced conical shape was made with a laser spot size of 6 $\mu$m with the depth of 1.5 mm.

Adjacent structures sometimes run into each other in dense structures, but the influence of the structure collapse on the muonium emission is unknown.
In a more macroscopic scale, the ablation on one side of the target induces the contraction of the ablated face, resulting in a warp. The warp is somewhat mitigated by having enough non-ablated margin around the ablated area in this study.

\begin{figure}[tbh] 
\begin{center} 
\includegraphics[width=0.8\columnwidth]{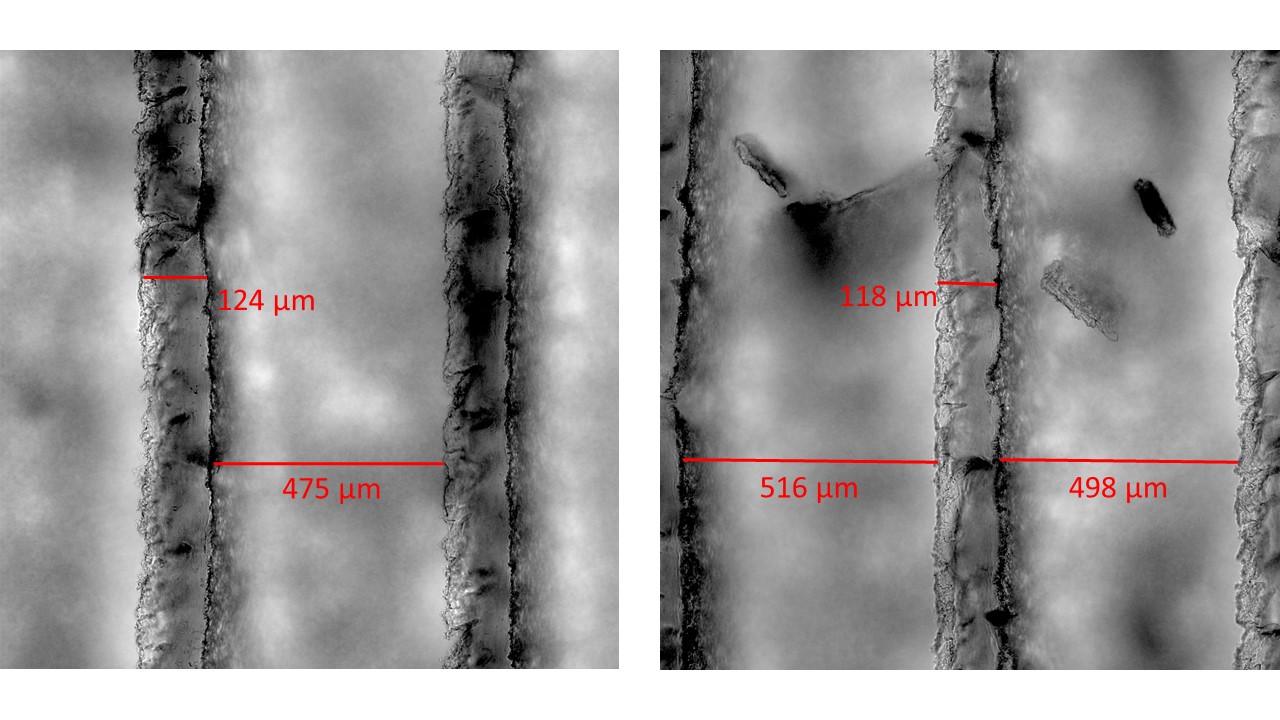} 
\caption{
Microscopic pictures of the aerogel target with channel ablation (S22), obtained using a laser-scanning confocal microscope as described in the text. The 120 $\mu$m-wide walls between the channels are in focus.
} 
\label{fig:channel_pattern} 
\end{center} 
\end{figure}

\begin{figure}[tbh] 
\begin{center} 
\includegraphics[width=0.4\columnwidth]{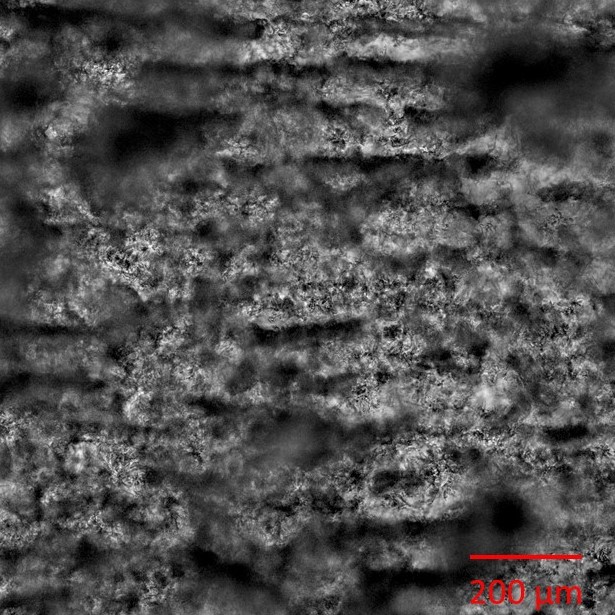} 
\includegraphics[width=0.4\columnwidth]{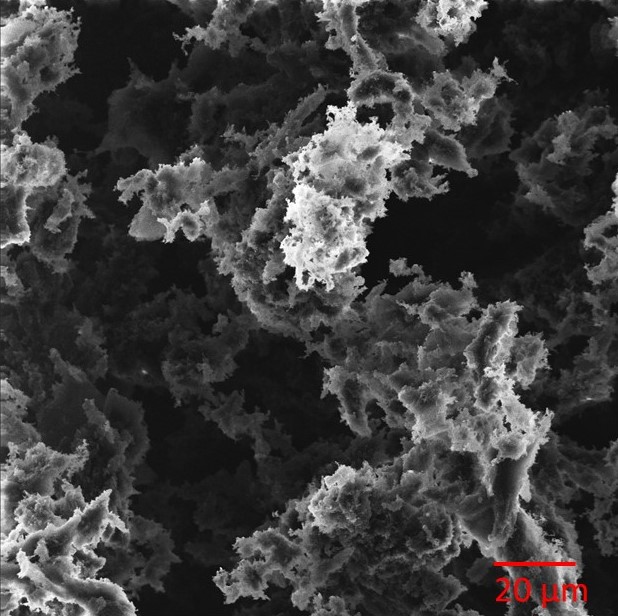}
\caption{
A microscopic picture of the aerogel target with continuous ablation (S14) (left) and details of the surface morphology (right). The left image is obtained using a laser-scanning confocal microscope as described in the text, showing the surface roughness over a large area of the target. The right image is acquired with Helium Ion microscope, showing the sub-micron features evidently.
} 
\label{fig:continuous_ablated_surface} 
\end{center} 
\end{figure}

Other structures can be processed by extending the hole ablation process.
A channel was made by continuous raster of the laser beam in one direction across the target. The speed of raster, typically around $1~{\rm mm/s}-2$ mm/s, determined the laser exposure and the resultant depth and width. For both wider channels and to achieve better definition of the walls separating each line, as many as 10 partially overlapping rasters were used. For example, the channels shown in Fig.~\ref{fig:channel_pattern} were produced by 10 line scans with 50 $\mu$m steps for each channel and 200 $\mu$m steps between channels. This resulted in about 500 $\mu$m-wide channels separated with about 120 $\mu$m-wide clean walls.
A V-groove was made in a similar fashion as the channel structures, however with using a higher numerical aperture lens to produce smaller spot size of 6 $\mu$m. Such a spot size was also utilized for the enhanced conical holes.
A target with a continuously ablated area was produced by the line raster of the laser beam over the entire ablation region, while each line was overlapped with adjacent lines at 50\% of the spot size. The pulse energy was set at very low level of about 30 $\mu$J, and scan speed of 2 mm/s. This resulted in a rough surface covered with microstructures with a depth of about 400 $\mu$m. Figure~\ref{fig:continuous_ablated_surface} shows the surface morphology of the continuously ablated target. The image on the right is acquired by Helium Ion microscope which shows pores and voids on the scale of $10~\mu{\rm m}-100$ $\mu$m. The presence of sub-micron features is also evident.

The previous study reported that the difference in the target weight measured before and after the laser ablation was only $5\%-10$\% of the expected difference, which was based on the ablation geometry and target density~\cite{PTEP2014}.
The same comparisons were carried out with a silica aerogel target used in this study. The obtained results are summarized in Table~\ref{tab:weight_loss}. The weight loss per ablated hole is found to be about 0.14 $\mu$g. This value is consistent with the expected weight loss based on the assumption of a conical shape for the holes. 

\begin{table}[!tbh]
\caption{
Weight loss due to laser ablation for an aerogel target with ablated holes (S18). The weights before and after ablation are measured with an electronic scale. The expected weight loss is evaluated assuming a conical shape with the intended diameter, pitch and depth, and has a range to account for the large variation of $\pm 250$ $\mu$m on the depth.
}
\label{tab:weight_loss}
\centering
\small
\begin{threeparttable}
\begin{tabular}{lc}
\hline\hline
\multicolumn{1}{c}{Item} & Value \\
\hline\hline
Weight before ablation & 525.45 mg\\
Weight after ablation & 523.09 mg\\
Weight loss due to ablation &  2.36 mg \\
Number of ablated holes & 16,680\\ 
Weight loss per ablated hole & 0.14 $\mu$g\\
Expected weight loss per ablated hole & $0.12~\mu{\rm g}-0.21$ $\mu$g\\
\hline\hline
\end{tabular}
\end{threeparttable}
\end{table}

\section{Measurements at TRIUMF}
\subsection{Beam condition and experimental setup}
\label{sec:setup}
\begin{figure}[htbp] 
\begin{center} 
\includegraphics[width=\columnwidth]{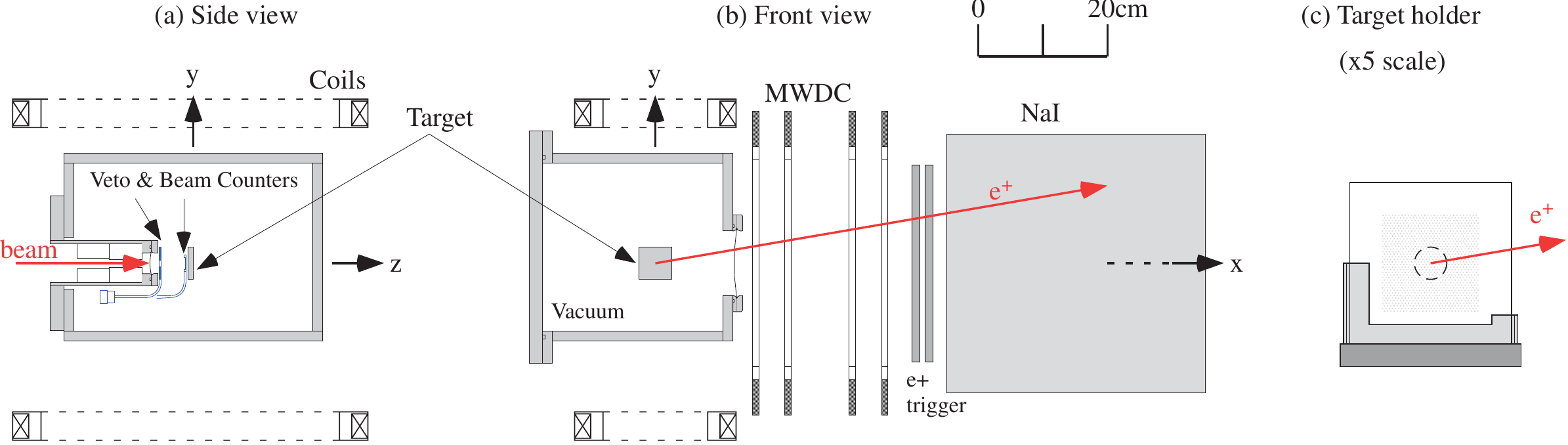} 
\caption{
 Schematics of the experimental setup used in the measurements at TRIUMF. Details are described in the text.
 Also shown is the enlarged figure of the target and the target holder seen from downstream. The shaded area around the center shows the ablated region and the circle shows the size of the hole of the veto counter.
} 
\label{fig:setup_chamber} 
\end{center} 
\end{figure}

In order to study the muonium emission from various structures on the target face, measurements were carried out at the M15 beamline of TRIUMF in 2017.
The beamline delivers a continuous beam of surface and subsurface~\cite{PTEP2013,Subsurface} muons. The typical muon rate at the beam counter, described below, was $1 \times 10^4$ s$^{-1}$ and the momentum spread was about 2\% (RMS) at a momentum of 23 MeV/$c$ with the momentum slit opening of 20 mm.
The experimental setup is shown in Fig.~\ref{fig:setup_chamber}. It was essentially the same as in our past studies~\cite{PTEP2013,PTEP2014} except for a few changes on the apparatus mentioned below.

The muonium production target was placed in the vacuum chamber operated at a vacuum level of around $10^{-4}~{\rm Pa}-10^{-3}$ Pa. This target chamber was connected to the beam port at the upstream wall and to the vacuum pumps at the downstream wall, and had an aluminized Mylar window, with the diameter of 100 mm and thickness of 100 $\mu$m, at a side wall next to the target.
The positrons decayed from muons, including those in muonium atoms, in the direction toward the Mylar window can enter the detection system with tracking capability outside of the target chamber.
The reconstructed positron tracks were extrapolated back to the vertical plane containing the beam axis to infer the muon decay positions. The obtained time and spacial distributions of the muon decay positions were used for the analysis.
Here, the detector coordinate system ($x$, $y$, $z$) is defined as shown in Fig. 5. The $z$-axis is along the beam center, $x$ towards the positron detection system, and $y$ vertically upward. The $z$-coordinate origin was the design position of the target downstream face.

The incoming subsurface muons passed through a series of two lead collimators held in the cylindrical housing at the upstream wall of the target chamber.
The collimator holes had the diameter of 16 mm and 12 mm, respectively, and both had the length of 50 mm.
The collimated muon beam entered the target chamber through a 50 $\mu$m-thick aluminized Mylar of a vacuum separator equipped at the downstream end of the housing. A plastic scintillator that has a 10 mm-diameter hole aligned to the last collimator hole was placed behind the housing end, and was used to veto events with unexpected beam trajectory. Another plastic scintillator was placed at $z= -12$ mm and was used to start a data taking cycle when fired. 
The thickness of this beam counter, 300 $\mu$m in the past studies, was increased to 380 $\mu$m in this study. The resulting increase in the deposited energy gave a better resolution in scintillation light detection and improved the discrimination of positrons from muons in the beam.

A muonium production target with a typical thickness of about 9 mm, or the calibration target of 0.1 mm-thick silica plate, was set on a target holder, which aligned the target downstream face to be $z=0$. The target holder was redesigned for this study to accommodate the larger target size than that in the past studies to mitigate the ablation warp as explained before. The target holder was made of poly(ether ether ketone) (PEEK) and held a target between fixed and removable covers on a base with gentle contact.
The covered area is limited to the 6 mm peripheral edges of large faces and the side faces, using only the lower half of the target. The detector side was further cut down to remove materials that may deteriorate the tracking resolution due to the multiple scattering of positrons.

The positron detection system consisted of four layers of multi-wire drift chamber (MWDC)~\cite{MWDC}, a pair of plastic scintillators and a NaI calorimeter.
The MWDC layers were used for positron tracking. A single layer consisted of a pair of wire planes tilted alternatively by $\pm 45$ degrees with respect to the $x-y$ plane.
Tracks reconstructed with good quality were required to pass through the aluminized Mylar window, centered at $z= 15$ mm, within a radius of 45 mm.
The track slope in the $x-z$ plane ($dz/dx$) was limited to be $|dz/dx| < 0.1$ to reduce the parallax between the true and inferred decay positions.
The vertical positions of muon decay points were required to be within $\pm$20 mm as in the past studies reflecting the beam size.
A pair of scintillators was used to issue the trigger to acquire the event and to determine the muon decay times.
The requirement on the muon decay time to be later than 6 ns removed the events of positrons that fired the beam counter and passed through the positron detection system. Such positrons can be produced, for instance, from the muons stopped around the beam counter.
The NaI calorimeter was used to reduce the influence of multiple scattering effect by selecting only high-energy positrons ($> 30$ MeV). This selection also contributed to higher sensitivity to spin polarization information. 

A pair of coils was installed above and below the target chamber in this study, providing a magnetic field for the spin precession study up to about 1.8 mT transverse to the beam direction.
Each coil, with the cross section of 35 mm $\times$ 25 mm, was wound on a frame that forms a rectangular shape with the inner dimensions of 462 mm and 162 mm in the longer and shorter directions, respectively. The two coils were positioned symmetrically to the detector coordinate origin with the gap of 440 mm. The field homogeneity is expected to be better than 0.3\% over the longitudinal range used in the analysis ($-20$ mm $< z < 40$ mm).

\subsection{Beam-momentum tuning}
\label{sec:beamMom}
The beam momentum was tuned to maximize the muonium emission from the target into vacuum. Since the stopping distribution is much broader than the muonium diffusion length, the maximum emission is effectively achieved by having the muonium population density, namely the Gaussian-like muon stopping distribution, peaked at or nearby the target emission face within the diffusion length. Half of the incident muons exit the target at this condition. This is determined by measuring the fraction of decay positrons originating in the target over a certain range of beam momentum.

\begin{figure}[!tbp]
  \begin{minipage}{0.5\hsize}
    \begin{center} 
      \includegraphics[width=\hsize, angle=0]{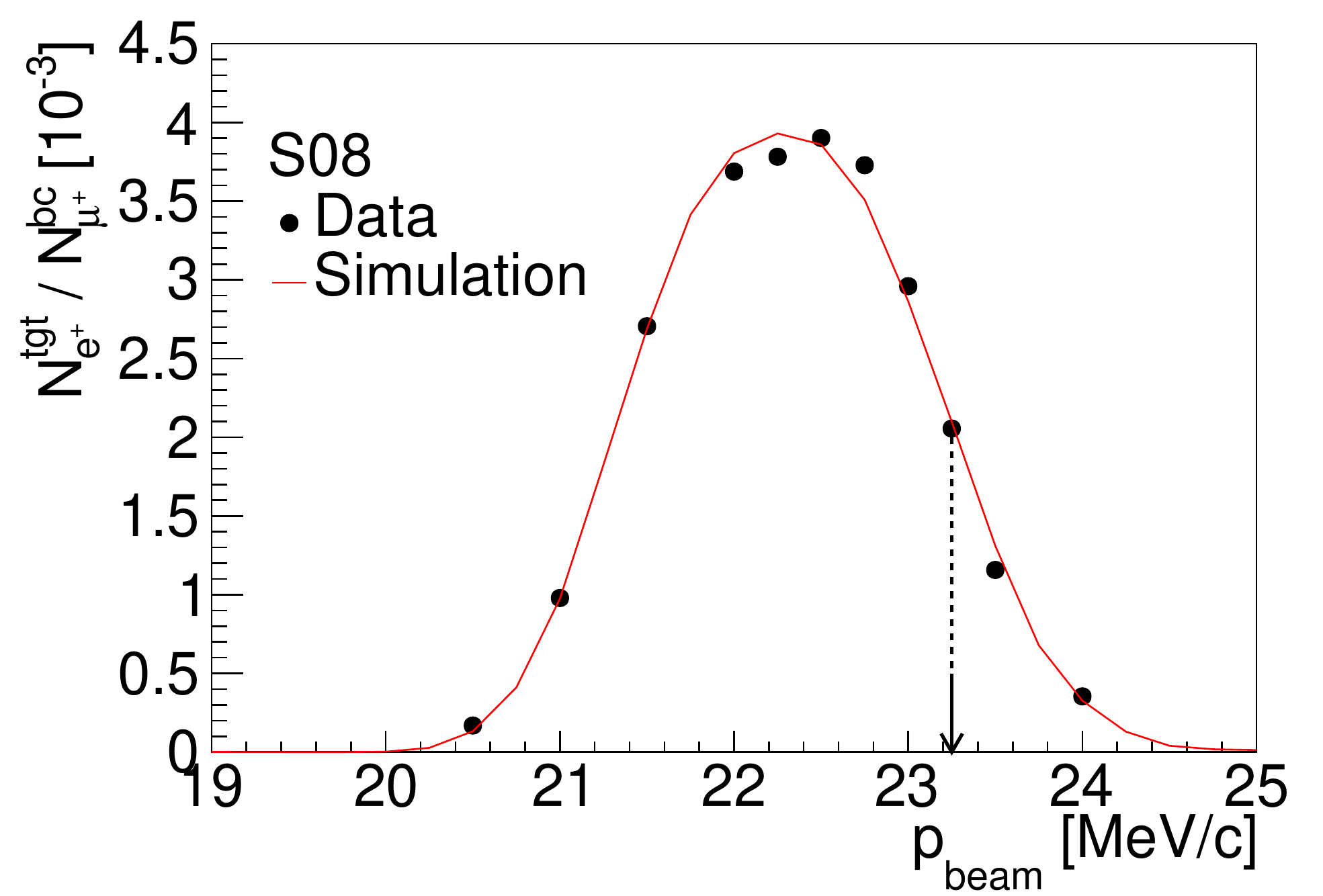}
    \end{center}
  \end{minipage}
  \begin{minipage}{0.5\hsize}
    \begin{center} 
      \includegraphics[width=\hsize, angle=0]{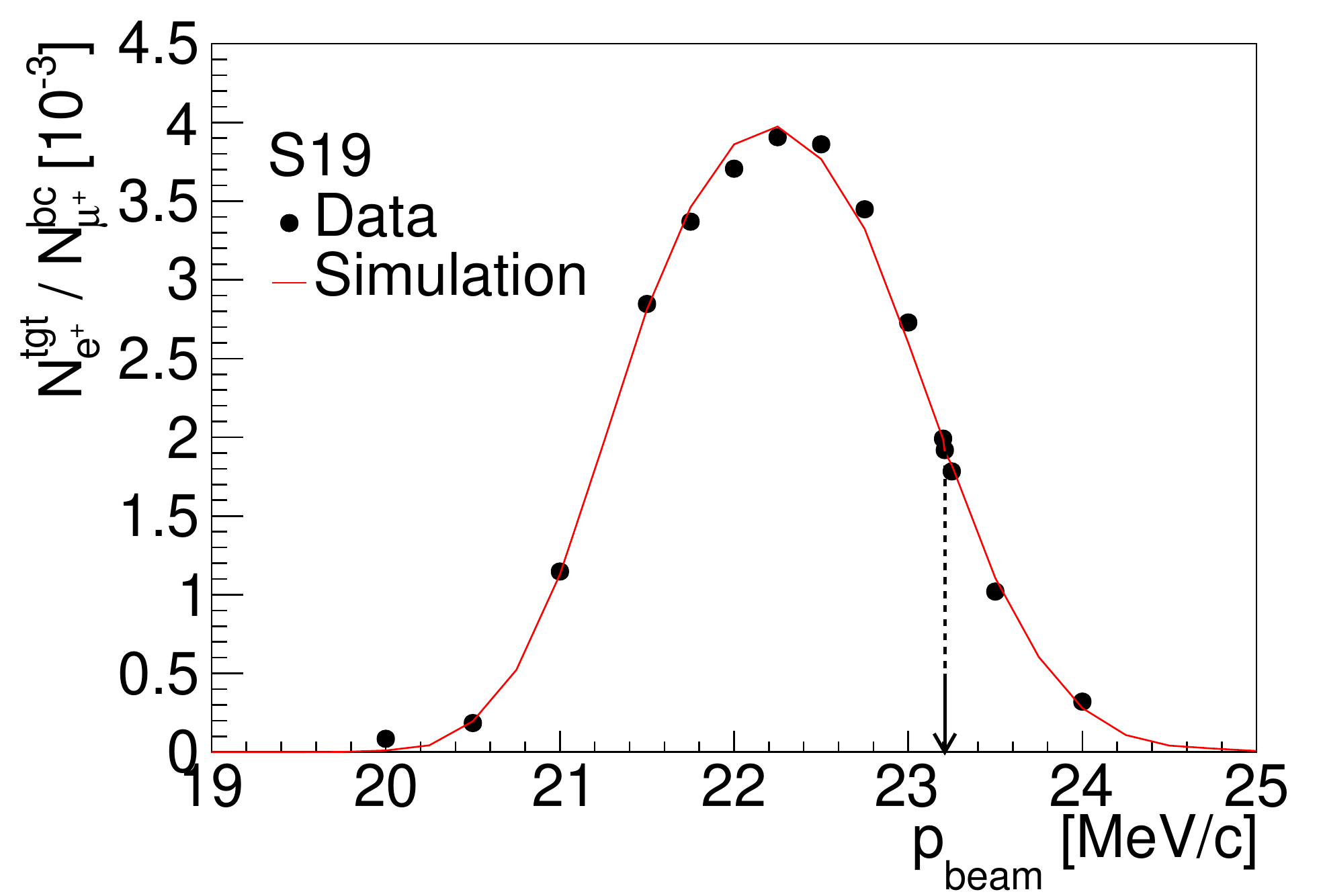}
    \end{center}
  \end{minipage}
  \begin{center}
    \caption{
      The variation of the number of positrons decayed from the target ($N_{e^+}^{\rm tgt}$) normalized by the number of incoming muons ($N_{\mu^+}^{\rm bc}$) as a function of the beam momentum ($p_{\rm beam}$) for the flat aerogel target (S08, left) and an aerogel target with hole ablation (S19, right). The black points show the measured positron yields. The red lines show their simulations described in the text. The arrows on the horizontal axes indicate the half-stop momenta at which the positron yields are half maximum.
     } 
  \label{fig:momscan_noabl} 
  \end{center} 
\end{figure} 

The corresponding fractions measured for the flat and an ablated aerogel targets, as well as their simulated ones, are shown in Fig.~\ref{fig:momscan_noabl} as a function of the beam momentum. The simulation curves were obtained from the muon stopping in the corresponding targets using the Geant4 tool kit~\cite{Geant4} with a Gaussian profile for the beam momentum. The best fit to the measured variation on the flat aerogel target was obtained with the momentum spread of 1.9\% (RMS). The positron yield reaches the maximum at the beam momentum of around 22.5 MeV/$c$. It decreases toward the lower beam momentum since more incident muons stop in the beam counter placed in front of the target, and similarly toward the higher momentum since more muons penetrate the target.
The beam momentum that makes the positron yield be half maximum in the higher-momentum side is the condition to maximize the muonium emission and is referred as ``half-stop'' momentum in this study. The reasonable agreement between the measurements and simulations supports the above understanding. The beam momentum is tuned for each target to account for the different structures on the target face before proceeding further studies.

\subsection{Muonium emission rates}
\label{sec:MuYield}
Incident surface muons slow down in the muonium production target, losing their kinetic energies by specific ionization energy loss and iteration of muonium formation/dissociation in a time scale of nano seconds or less~\cite{Thermalization}.
Given the half-stop momentum condition, nearly half of them penetrate the target during this process and decay downstream of the target or hit the vacuum chamber wall.
Thermalization of the rest of stopping muons to the temperature of the material involves elastic and inelastic collisions of both muons and muonium atoms with the lattice and atoms of the medium; some thermalized muons also end up in muonium atoms after capturing radiolysis electrons~\cite{DMF}. The empirical fraction of muons forming muonium atoms in the target is to be discussed in Sec.~\ref{sec:MuFrac}.
Neutral muonium atoms can diffuse inside the target through random collisions until they arrive at the downstream surface of the target, where some emerge into vacuum and are emitted downstream with a room temperature Maxwell-Boltzmann velocity distribution.
Muons in the emitted muonium atoms mostly decay nearby the target and are indistinguishable from those decaying inside the target, referred to as ``target decays'', because of the limited spatial resolution of the positron track extrapolation.
A fraction of the emitted muonium atoms continue on trajectories away from the target, giving a time-of-flight (TOF) evolution of the decay point distribution in vacuum~\cite{PTEP2013}.
These muonium atoms in vacuum were used to evaluate the muonium emission rate.

\begin{figure}[!htbp]
  \begin{minipage}{0.5\hsize}
    \begin{center} 
      \includegraphics[width=\hsize, angle=0]{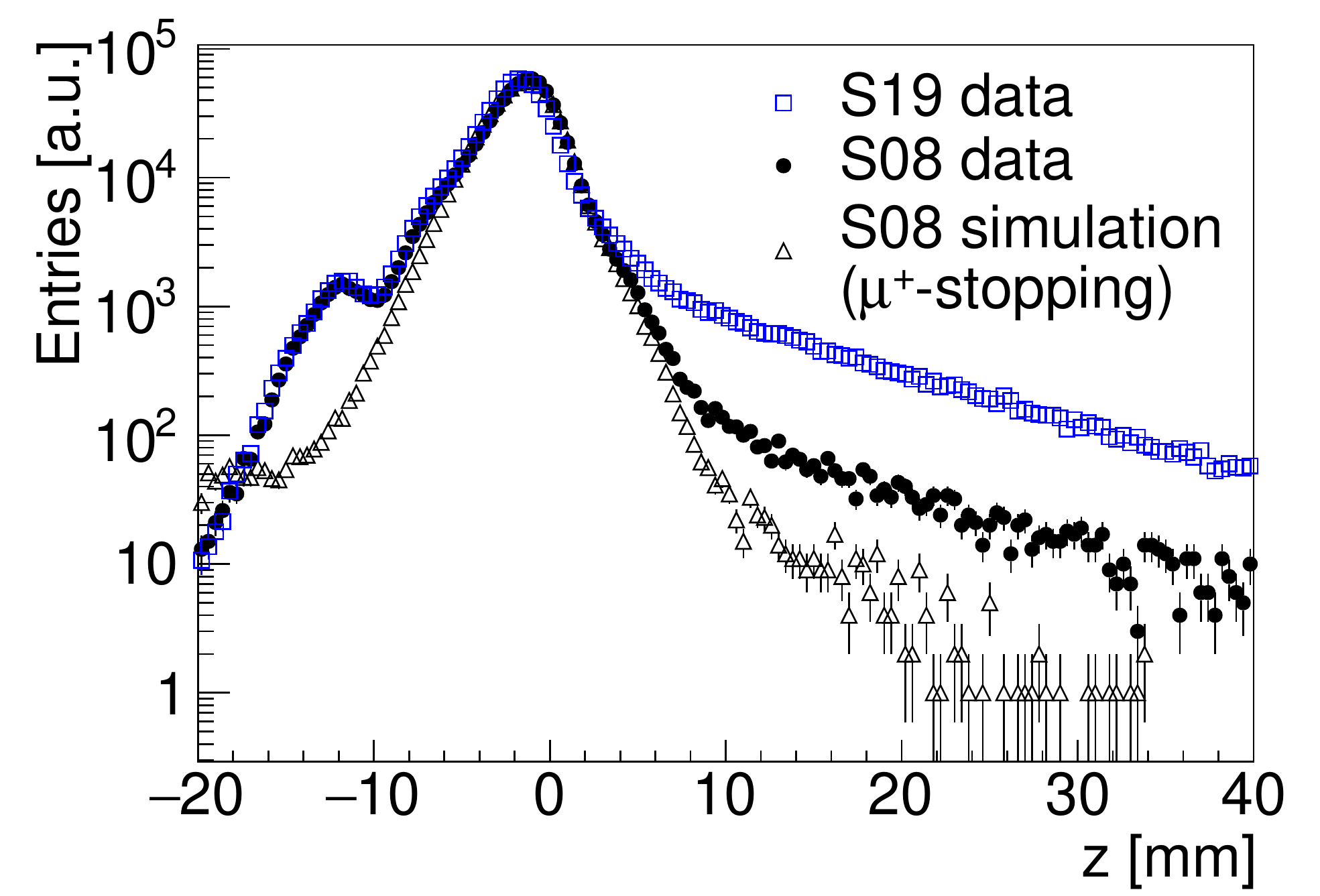}
    \end{center}
  \end{minipage}
  \begin{minipage}{0.5\hsize}
    \begin{center} 
      \includegraphics[width=\hsize, angle=0]{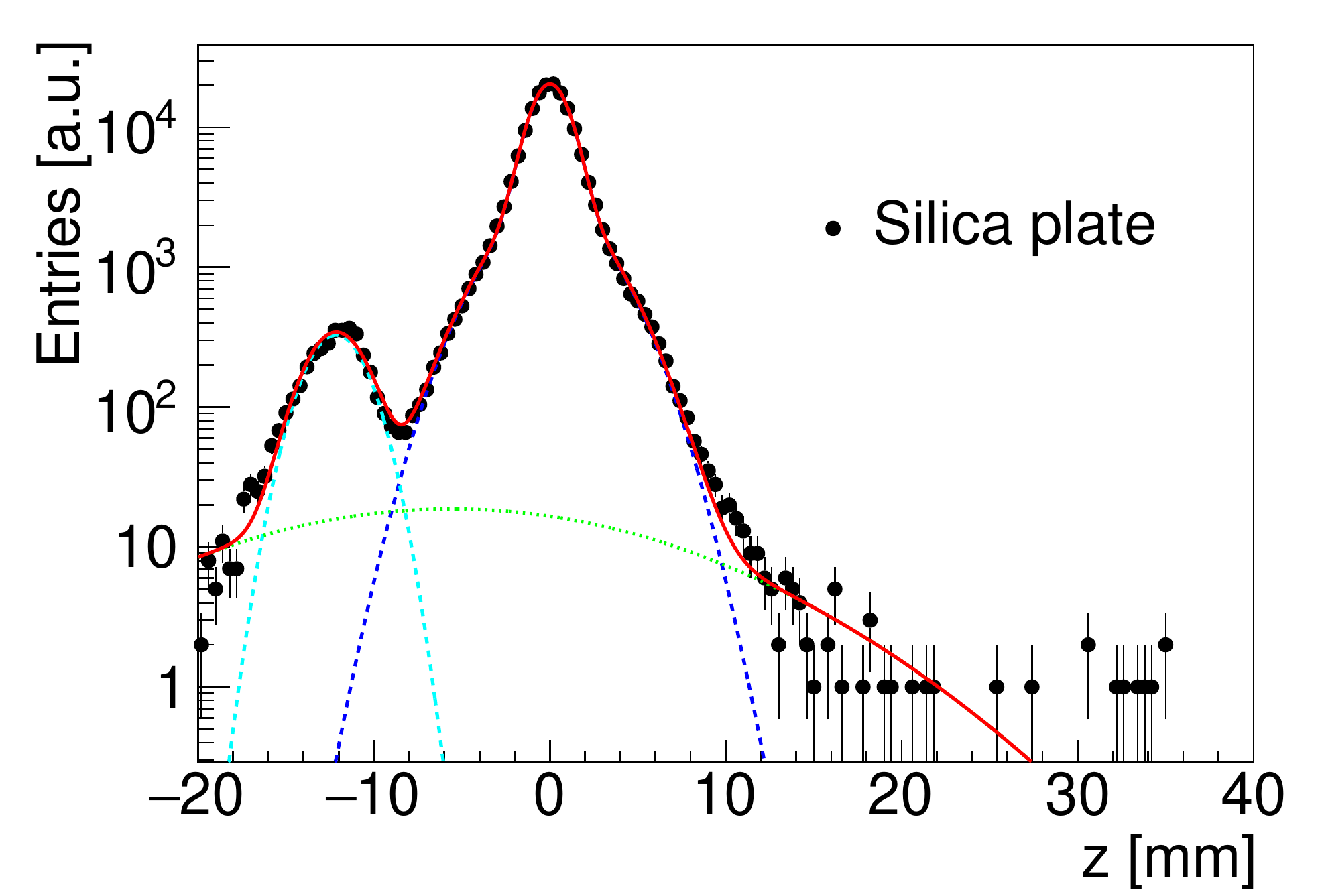}
    \end{center}
  \end{minipage}
  \begin{center}
    \caption{
        The $z$-position distributions of extrapolated positron tracks in the $y-z$ plane are shown for the data taken with aerogel targets (left) and silica plate (right). The aerogel targets used are the flat (S08, black closed circle) and a hole-ablated (S19, blue open square) ones, and the distribution for the latter is scaled to have the same maximum height with that for the former.
        The $z$-positions of simulated muon decays inside the flat aerogel target (black open circle), convoluted with a resolution function described in the text, are also shown in the left.
        The resolution of positron track extrapolation is evaluated from the fit on the distribution in the silica plate data (red solid line) including the components for muonium and muon decays in the silica plate (blue dashed line) and beam counter (light-blue dashed line) as well as for outlier events (green dashed line) as described in the text.
     } 
  \label{fig:zdist_triumf} 
  \end{center} 
\end{figure} 

Figure~\ref{fig:zdist_triumf} shows the distributions for the $z$-positions of muon decay points ($z_{\rm dec}$) in the data taken with aerogel targets (left) and silica plate (right).
The $z_{\rm dec}$ distributions for the aerogel targets show a clear enhancement of downstream muonium decays in vacuum for the hole-ablated target compared to the flat one as seen in the previous study~\cite{PTEP2014}.
The corresponding distribution of target decays only ($z_{\rm dec}^{\rm tgt}$) was simulated for the flat aerogel target, discussed later, and is shown in the same figure for the case with no muonium emission.
The simulated distribution in the positive $z$ region rapidly drops to the level of an order of magnitude smaller than that for the measured data at $z=10$ mm, while the measured data distribution extends further away from the target emission face because of the emitted thermal muonium atoms.
Therefore, the muonium emission rate can be measured in the region of $z > 10$ mm with significantly suppressed background contribution and is defined as
\begin{equation}
R_{\rm Mu}^{\rm vac} = \frac{N_{e^+}^{\rm vac}}{N_{e^+}^{\rm all}},
\end{equation}
where the $N_{e^+}^{\rm vac}$ and $N_{e^+}^{\rm all}$ denote the number of positron tracks extrapolated to the vacuum region (10 mm $< z <$ 40 mm) and all region ($-20$ mm $< z <$ 40 mm), respectively, in the $y-z$ plane.

The $z_{\rm dec}^{\rm tgt}$ distribution for the flat aerogel target was produced taking the simulated decay points of stopping muons in the target, described in Sec.~\ref{sec:beamMom}, and no decays in the beam counter were involved.
The simulated $z$-position distribution was smeared by convoluting with the resolution shape taken from the corresponding $z_{\rm dec}^{\rm tgt}$ distribution in the silica plate data, shown in Fig.~\ref{fig:zdist_triumf} (right), which involves no muonium atoms decaying outside the silica plate because of the high density and is dominated by the extrapolation resolution because of the small thickness; the extrapolation resolution was evaluated to be about 2.5 mm (FWHM) from this distribution.
The $z_{\rm dec}^{\rm tgt}$ distribution in the silica plate data was modeled with two Gaussian functions for the target decays, sharing the mean, and single Gaussian functions for the beam counter decays and outlier events.
All Gaussian parameters, other than the one shared, were allowed to vary in the fit. The obtained shape without the component for the beam counter decays was taken as the resolution function.
The peak position and height of smeared simulation were adjusted to fit the $z_{\rm dec}$ distribution in the flat aerogel data over the target decay dominant range ($-8\ {\rm mm} < z < 8\ {\rm mm}$).
The smeared simulation of $z_{\rm dec}^{\rm tgt}$ distribution was used to evaluate the background contributions of target decays and outlier events in vacuum.
It was also used to estimate imperfect positioning of targets discussed later.

\begin{table}[!hbtp]
\caption{
Summary of the muonium emission rates in vacuum ($R_{\rm Mu}^{\rm vac}$) and their statistical uncertainties measured over the ablation structures listed in Table~\ref{tab_targets} at the TRIUMF M15 beamline.
The groove-ablated targets are labeled with ``H'' and ``V'', indicating the horizontal and vertical orientations, respectively, of groove configuration in the measurements.
The emission rates measured in the previous study~\cite{PTEP2014} are also listed for comparison.
Note that the silica plate (SP) has no muonium emission and its rate represents the background contributions only as described in the text.
}
\label{tab:yields_summary_aerogel}
\centering
\small
\begin{tabular}{ccrcccr}
\hline\hline
Ablation & Target & \multicolumn{1}{c}{$R_{\rm Mu}^{\rm vac}$ [\textperthousand]} & &
Ablation & Target & \multicolumn{1}{c}{$R_{\rm Mu}^{\rm vac}$ [\textperthousand]} \rule[-1mm]{0mm}{4mm}\\
\hline\hline
Flat  & SP  & $0.75 \pm 0.07$ \\ 
\hline
Flat & S08 & $2.92 \pm 0.06$ & & \multirow{5}{*}{Grooves} & S21H & $26.3 \pm 0.3$\ \,  \\
\cdashline{1-1}
\multirow{13}{*}{Holes} & S01 & $22.1 \pm 0.2$\ \,  & & & S21V & $24.3 \pm 0.3$\ \,  \\
& S04 & $19.0 \pm 0.3$\ \,  & & & S22H & $23.7 \pm 0.3$\ \,  \\
& S05 & $10.6 \pm 0.2$\ \,  & & & S22V & $25.9 \pm 0.3$\ \,  \\
& S09 & $21.8 \pm 0.2$\ \,  & & & S24H & $22.2 \pm 0.3$\ \,  \\
\cdashline{5-5}
& S10 & $18.8 \pm 0.1$\ \,  & & Continuous & S14 & $15.2 \pm 0.1$\ \,  \\
\cline{5-7}
& S11 & $10.9 \pm 0.1$\ \,  & & Flat  & S15 & $ 4.27 \pm 0.14$ \\
& S12 & $ 5.72 \pm 0.15$ & & Holes & S16 & $12.2 \pm 0.1$\ \,  \\
\cdashline{5-5}
& S13 & $ 8.23 \pm 0.10$ & & Flat  & S17 & $ 3.21 \pm 0.11$ \\
& S18 & $27.6 \pm 0.2$\ \,  & & Holes & S25 & $ 8.56 \pm 0.23$ \\
\cline{5-7}
& S19 & $26.9 \pm 0.1$\ \,  & & Flat  & O2013-1 & $ 3.72 \pm 0.11$ \\
\cdashline{5-5}
& S20 & $22.0 \pm 0.3$\ \,  & & \multirow{3}{*}{Holes} & O2013-2 & $16.0\pm 0.2$\ \,  \\
& S2013-1 & $17.2 \pm 0.2$\ \,  & & & O2013-3 & $20.9 \pm 0.7$\ \,  \\
& S23 & $14.6 \pm 0.2$\ \,  & & & O2013-4 & $30.5 \pm 0.3$\ \,  \\
\hline\hline
\end{tabular}
\end{table}

The measured muonium emission rates are summarized in Table~\ref{tab:yields_summary_aerogel}.
The background contributions from the target decays and outlier events evaluated using the smeared simulation for each target were subtracted from the observed entries in the vacuum region.
Each of the smeared $z_{\rm dec}^{\rm tgt}$ distributions for ablated targets is based on the muon stopping distribution for the flat target of corresponding target material, weighting the distribution near the surface according to the material loss expected from its ablation structure.
The measured rates are in the range 5\textperthousand$-$28\textperthousand\ and below 5\textperthousand\ for the ablated and flat targets, respectively.
Enhancement of muonium emission for all the ablated targets to the flat targets of corresponding materials is confirmed, including the PMSQ targets which have higher densities than those of the silica aerogel targets.
Some of the targets show the emission rates close to the highest one measured in the previous study~\cite{PTEP2014}, which can be applicable to an early phase of the J-PARC muon $g$-2/EDM experiment~\cite{E34}.

The dominant systematic uncertainties on the emission rates come from the imperfect target positioning and the background modeling.
The target emission faces were slightly tilted to and/or displaced from the designed plane at the coordinate origin because of the natural aerogel shape and the gentle fixation on the target holder.
The tilt of each target was evaluated fitting a line to the peak position variation in the $z_{\rm dec}$ distribution over the central 15 mm range in y-coordinate.
The displacement of each target was evaluated from the difference between the $z$-intercept of the tilt line in the $y-z$ plane and the peak $z$-position of the smeared simulation with no displacement.
The muonium emission rate was evaluated again with the background subtraction in the regions redefined with the tilt and displacement.
The resulting relative changes from the nominal results were estimated to be $-6$\% to $+12$\% over the targets as the uncertainties on the imperfect target positioning. 
The uncertainty on the background modeling is estimated using the silica plate data as an alternative background model since it has no muonium emission.
The event rate in the nominal vacuum region of the silica plate data is subtracted from that of each target data, instead of the background subtraction mentioned above. The resulting relative changes from the nominal results ranged in $-6$\% to $+1$\%.
Adding the uncertainties on the imperfect target positioning and background modeling in quadrature, the relative systematic uncertainties on the emission rates were estimated to be $-7\%$ to $+12$\%.

\begin{figure}[!tbp] 
\begin{center} 
\includegraphics[width=0.7\hsize, angle=0]{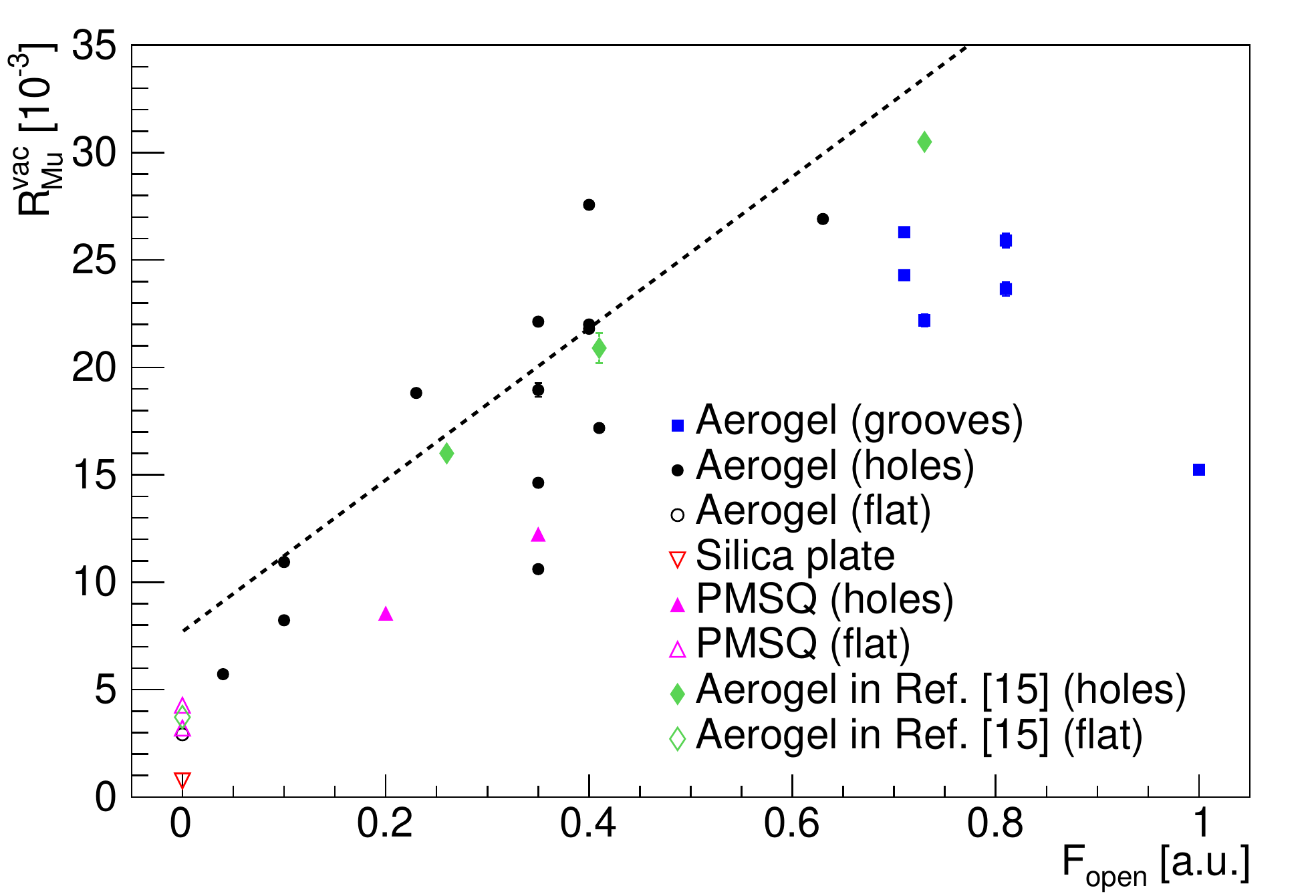} 
\caption{
The dependence of the muonium emission rate ($R_{\rm Mu}^{\rm vac}$) on the opening fraction of the ablation region ($F_{\rm open}$).
The error bars show the statistical uncertainties only.
The closed circle (black) and square (blue) represent the aerogel targets with hole- and groove-ablation, respectively, and the upward-triangle (magenta) represents the PMSQ-gel targets with hole-ablation. The aerogel targets with hole-ablation measured in Ref.~\cite{PTEP2014}, the closed diamond (green), are also shown for comparison. The open symbols represent the non-ablated targets of the corresponding closed symbols, except the downward-triangle which represents the silica plate.
The dashed line is drawn as an eye-guide of the dependence seen over the aerogel targets with hole-ablation used in this study.
 } 
\label{fig:corr_area} 
\end{center} 

\end{figure} 
\begin{figure}[!hbtp]
  \begin{minipage}{0.5\hsize}
    \begin{center} 
      \includegraphics[width=\hsize, angle=0]{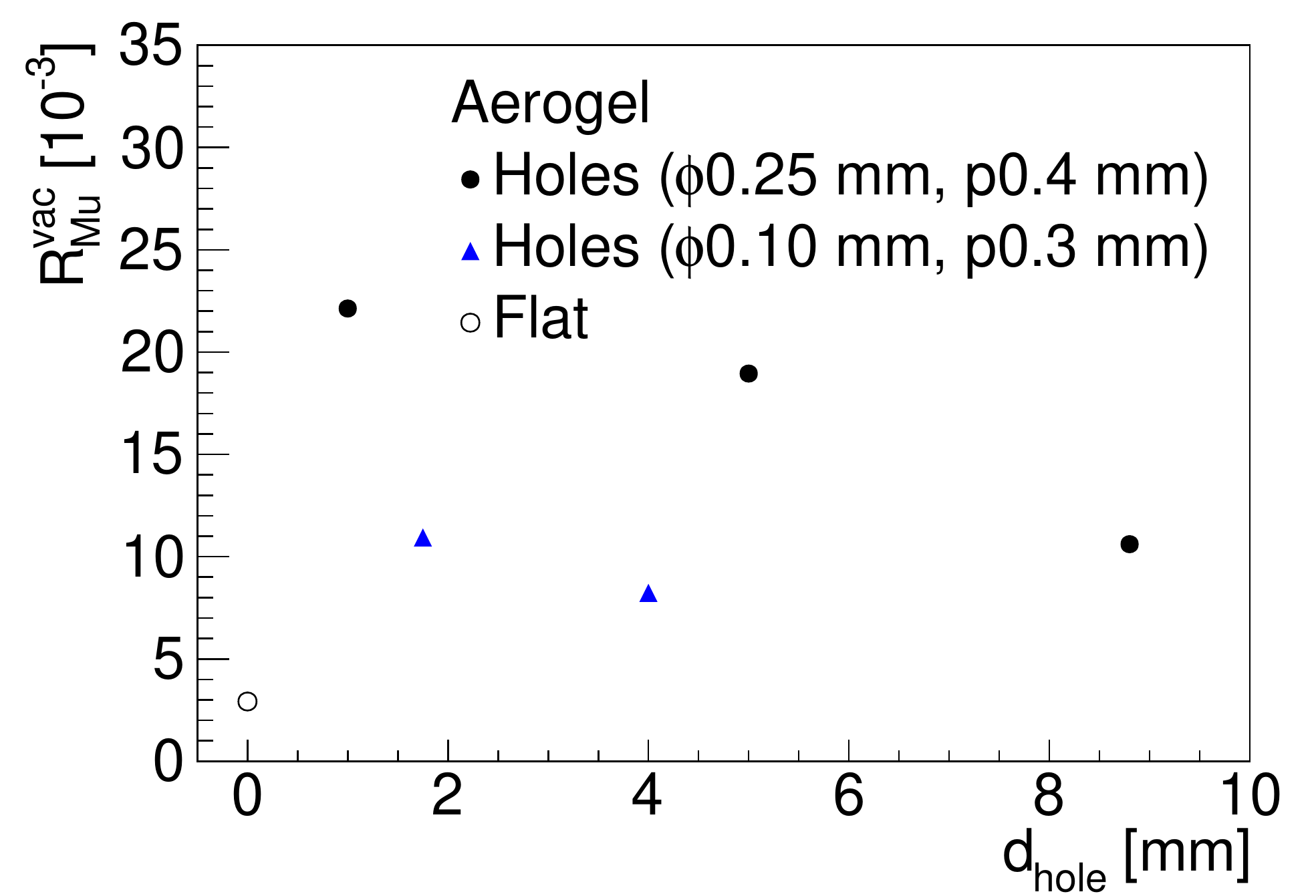}
    \end{center}
  \end{minipage}
  \begin{minipage}{0.5\hsize}
    \begin{center} 
      \includegraphics[width=\hsize, angle=0]{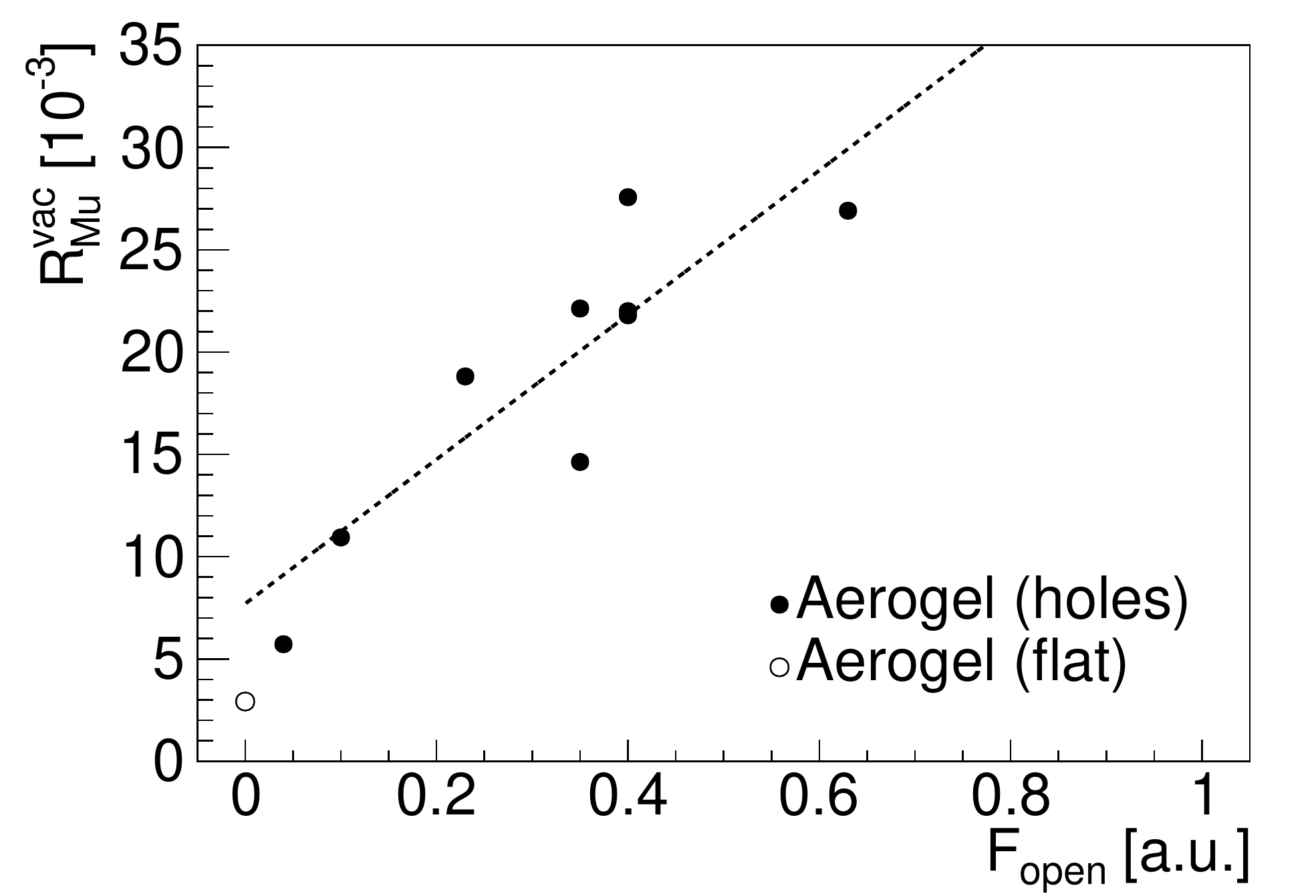}
    \end{center}
  \end{minipage}
  \begin{center}
    \caption{
        (Left) The dependence of the muonium emission rate in vacuum ($R_{\rm Mu}^{\rm vac}$) on the hole-ablation depth ($d_{\rm hole}$) for the sets of aerogel targets with the same diameter and pitch.
        The error bars show the statistical uncertainties only.
        The close circle (black) represents the aerogel targets with the diameter of 0.25 mm at the pitch of 0.4 mm, and the triangle (blue) represents the ones with the diameter of 0.10 mm at the pitch of 0.3 mm.
        The aerogel target with no ablation, open circle, is also shown as a reference.
        (Right) The emission rate dependence on the opening fraction of the ablation region ($F_{\rm open}$) for the hole-ablated aerogel targets with the ablation depth of 2 mm or less.
     } 
  \label{fig:corr_holes} 
  \end{center} 
\end{figure} 

The dependence of muonium emission rate on the ablation structure parameters was studied to understand the emission mechanism.
Naively the muonium emission rate is expected to depend on the surface area of ablation, i.e. the conical area for a hole,
since the laser ablation gives a macroscopic and microscopic structures to enhance muonium atoms diffuse out of the target. 
No obvious correlation was, however, seen between the emission rates and such area.
Surveying the dependence on the other combinations of the ablation structure parameters, it is found that the opening fraction of the ablation region ($F_{\rm open}$) shows such dependence for the silica aerogel targets with hole ablation as shown in Fig.~\ref{fig:corr_area}.
The dependence also applies to the ones used in Ref.~\cite{PTEP2014}.
The PMSQ-gel targets and the aerogel targets with groove ablation show deviations from the eye-guide with slightly lower emission rates. The rate for the continuously ablated target ($F_{\rm open} = 1$) is even lower. Further study is required to clarify this tendency.

Most of the ablated targets have the ablation depth of 2 mm or less, and no clear dependence of the muonium emission rate is seen on the ablation depth. The contribution of the ablation depth was specifically inspected by comparing the emission rates for the hole-ablated targets with different depths, keeping the other parameters the same. There are two sets of hole-ablated targets which meet this condition. The results indicate a slightly negative correlation as shown in Fig.~\ref{fig:corr_holes} (left). The tendencies suggest an optimal depth to be less than 2 mm to maximize the muonium emission. A further study is necessary to understand the underlying mechanism and to find the optimal depth. Clearer rate dependence on the opening fraction can be seen in Fig.~\ref{fig:corr_holes} (right) by selecting the data only from the hole-ablated aerogel targets with the ablation depth of 2 mm or less.

\begin{figure}[htbp] 
\begin{center} 
\includegraphics[width=0.7\hsize, angle=0]{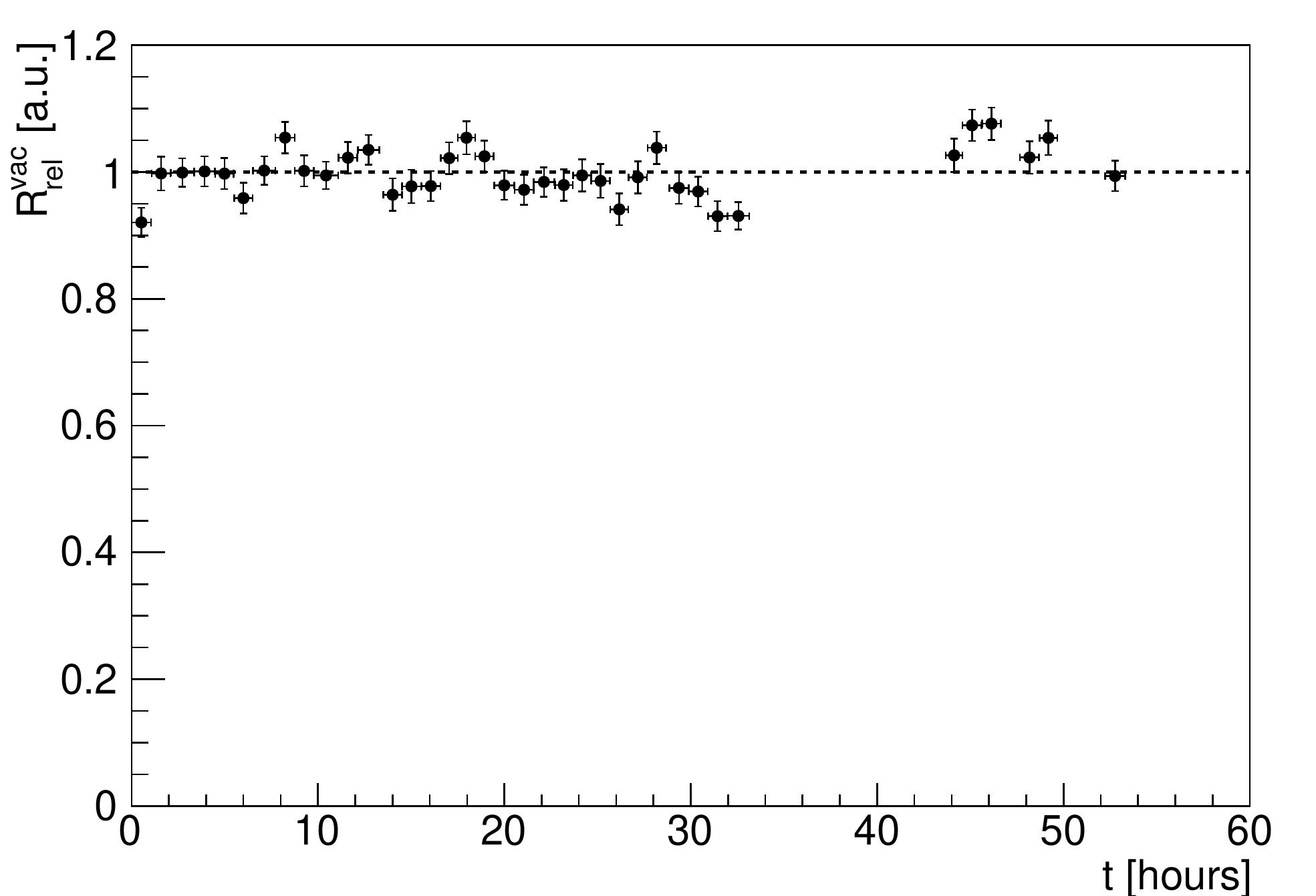} 
\caption{
    The relative variation of the muonium emission rate in vacuum ($R_{\rm rel}^{\rm vac}$) as a function of time ($t$). The rates are measured with the target S19 and normalized to its total emission rate listed in Table~\ref{tab:yields_summary_aerogel}. The vertical error bars show the statistical uncertainties only. The horizontal error bars indicate the corresponding data taking duration. The dashed line at $R_{\rm rel}^{\rm vac} = 1$ is given for an eye guide. The measurements were continued for 53.3 hours except for the 10.5 hours of accelerator maintenance at around $t = 40$.
} 
\label{fig:aging_s19} 
\end{center} 
\end{figure}

A long-term use with constant performance is desired for the muonium production target, but a long-term exposure to the high-vacuum environment or muon beam irradiation may degrade the muonium emission rate due to the possible material modification or charge-up effect, respectively.
The relative variation of the muonium emission rate was evaluated without the background subtraction over more than 50 hours, though including 10.5 hours of beam-off due to the accelerator maintenance, to address this issue and shown in Fig.~\ref{fig:aging_s19}. The emission rate was found to be constant during this period.
No physical damage from this extended exposure to the high vacuum environment was observed.

\subsection{Muonium formation}
\subsubsection{Muonium formation fraction}
\label{sec:MuFrac}

All $\mu^+$SR measurements refer to and rely upon 
the angular {\sl asymmetry\/} of muon decay.  
While the theoretical asymmetry averaged over all positron energies is 1/3, 
the asymmetry at the highest positron energy is 1 
(assuming perfectly polarized muons) 
and the average over detector solid angle is somewhat smaller.  
When the positron detector rejects low energy positrons 
and subtends a substantial solid angle, 
the experimental asymmetry is hard to predict.  
In any given experiment, therefore, 
the maximum asymmetry possible, $A_0$, 
is determined empirically from a spectrum obtained 
using a sample that preserves the initial muon polarization; 
a pure silver or aluminum sample is generally a good choice.  
Then all the experimental idiosyncracies are assumed to be 
identical in the same apparatus under the same conditions 
for any other sample with the same density and geometry.  
We determined $A_0 = 0.474 \pm 0.004$ in an aluminum plate 
with a magnetic field of $\approx 1.92$~mT, in which the applied and ambient fields were combined.

When polarized positive muons form muonium from unpolarized electrons, 
half the initial states are in the $\vert \Uparrow \uparrow \rangle$ eigenstate 
and half are in the $\vert \Uparrow \downarrow \rangle$ superposition state 
(where $\Uparrow$ represents the muon spin direction and 
$\updownarrow$ that of the electron), 
in which the muon polarization oscillates between $+1$ and $-1$ 
at the hyperfine frequency 
(in vacuum, $\nu_0 = 4463.302$ MHz~\cite{Mariam1982}), 
which is unresolved by most experiments.  
Thus the muonium polarization appears to be halved.  
Assuming that no ``prompt'' depolarization takes place 
during the stopping of the surface muons, 
the fraction $f_{\rm Mu}$ of muons forming polarized muonium 
in the target can therefore be determined by comparing 
{\sl twice\/} the measured muonium asymmetry $A_{\rm Mu}$ 
with the maximum available asymmetry $A_0$: 
\begin{equation}
 f_{\rm Mu} = \frac{2A_{\rm Mu}}{A_0} \; .
\label{eq:MuFraction}
\end{equation}

\begin{figure}[htbp] 
\begin{center} 
\includegraphics[width=0.65\hsize,angle=0]{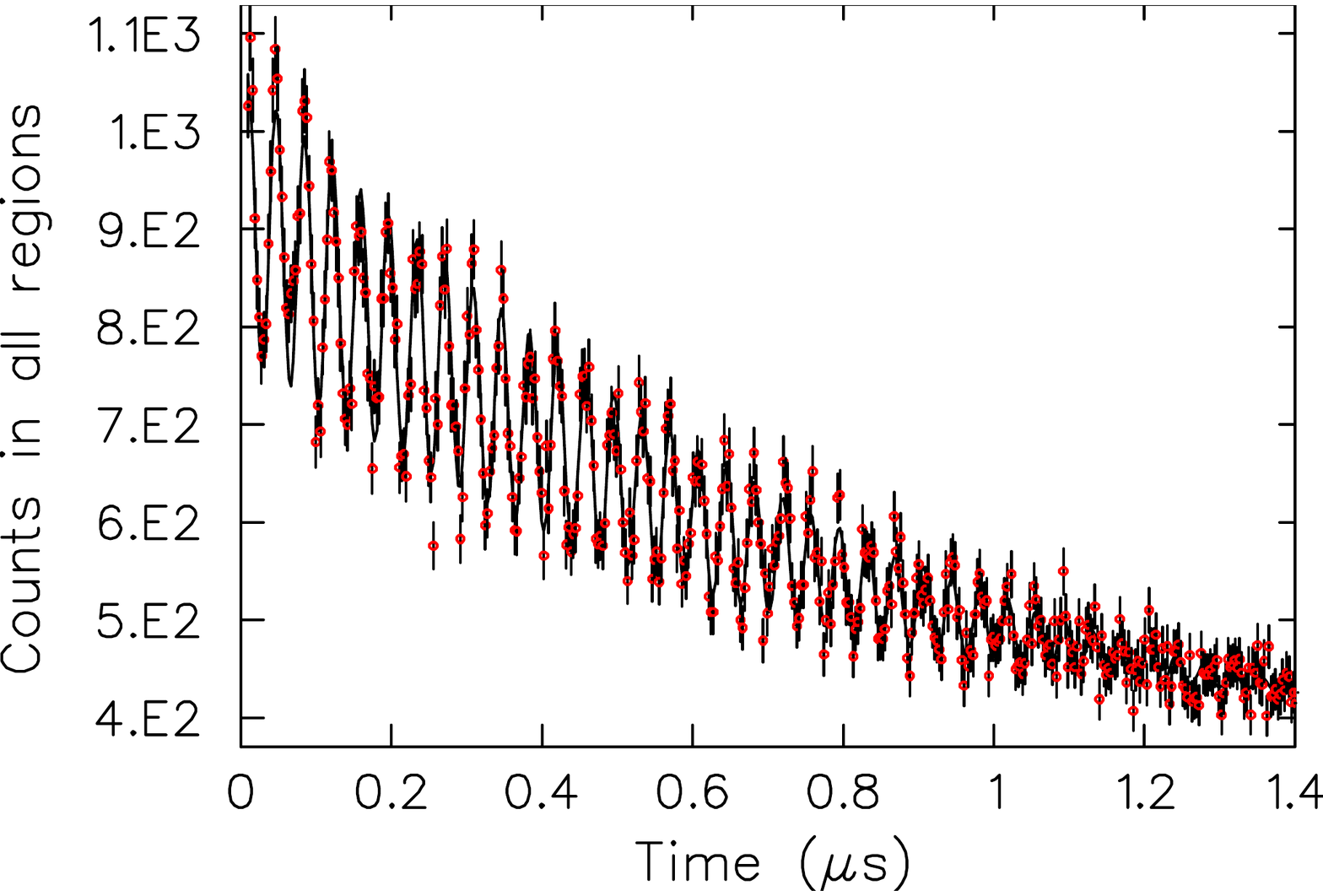} \\
\vspace{-3 cm}
\includegraphics[width=0.65\hsize,angle=0]{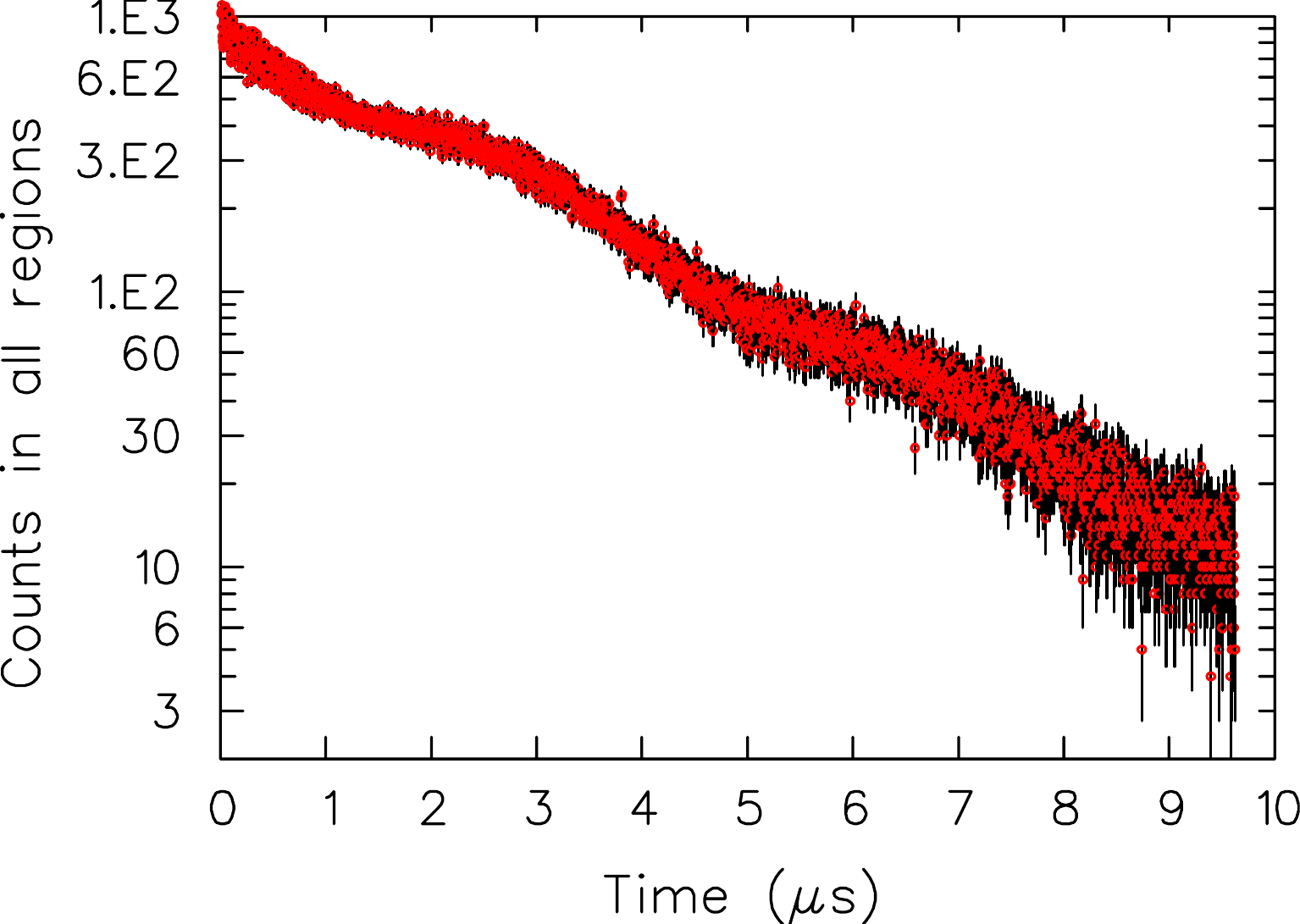} 
\caption{
 Fit of Eq.~(\ref{eq:NofT}) to the $\mu$SR signal from 
 the entire $z$ region ($-20\ {\rm mm} < z < 40\ {\rm mm}$) 
 in a sum of all data on silica aerogel targets 
 in a magnetic field of $\approx 1.92$~mT. 
} 
\label{fig:Ntot} 
\end{center} 
\end{figure}

Measuring the muonium asymmetry in aerogel targets 
using the same magnetic field ($\approx 1.92$~mT) 
requires fitting {\sl two\/} muonium frequencies, 
$\nu_{12}$ and $\nu_{23}$, corresponding to transitions 
between levels 1 \& 2 or 2 \& 3 of the muonium Breit-Rabi energy level diagram.  
In such a modest field, these two signals 
are of essentially the same amplitude 
but are {\sl split\/} by $\pm \Delta\nu_{\rm HF} = \nu_{\rm Mu}^2 / \nu_0$ 
where $\nu_{\rm Mu}$ is their average.  
This splitting is manifest in the spectrum shown in Fig.~\ref{fig:Ntot} 
as a ``beat'' pattern with the first node at about 1.56~$\mu$s.  

In the data comprising the spectrum in Fig.~\ref{fig:Ntot}, 
there are 12.47(1) three-nanosecond time bins per period of muonium precession, 
which makes display tedious 
but should provide a reliable determination of 
both muonium and diamagnetic asymmetries 
by a fit to the usual form of a transverse field $\mu^+$SR time histogram:  
\begin{eqnarray}
 N(t) &=& BG + N_0 e^{-t/\tau_\mu}[1 +  
  A_{\rm D} e^{-\lambda_{\rm D} t} \cos(2\pi\nu_{\rm D} t + \phi_{\rm D}) \cr
 &+& (A_{\rm Mu}/2) e^{-\lambda_{\rm Mu} t} \left\{ 
     \cos(2\pi\nu_{12} t + \phi_{\rm Mu}) + 
     \cos(2\pi\nu_{23} t + \phi_{\rm Mu}) \right\} ] \; ,
\label{eq:NofT}
\end{eqnarray}
where $BG$ (a time-independent background) is undetectable, 
$N_0 = (919.3 \pm 1.2)$ counts per 3-ns bin is the overall normalization, 
$\tau_\mu = 2.19703~\mu$s is the (fixed) muon lifetime, 
$A_{\rm Mu} = 0.155 \pm 0.004$ is the initial muonium asymmetry, 
$\lambda_{\rm Mu} = (0.28 \pm 0.02) \; \mu{\rm s}^{-1}$ 
is the muonium relaxation rate, 
$\nu_{12} = \nu_{\rm Mu} + \Delta\nu_{\rm HF}$ 
and $\nu_{23} = \nu_{\rm Mu} - \Delta\nu_{\rm HF}$ 
are the two frequencies of muonium precession 
split by twice $\Delta\nu_{\rm HF} = (0.1682 \pm 0.0027)$ MHz, 
whose average $\nu_{\rm Mu} = (26.7441 \pm 0.0027)$ MHz 
$= [\gamma_{\rm Mu}/(2\pi)] B$ 
[$\gamma_{\rm Mu} = 8.7617 \times 10^{10}$~rad~s$^{-1}$~T$^{-1}$] 
is the nominal muonium precession frequency, and 
$\phi_{\rm Mu} = (-91.2 \pm 0.8)$ degrees is 
the initial phase of the muonium precession; 
$A_{\rm D} = 0.165 \pm  0.002$ and 
$\lambda_{\rm D} = (0.049 \pm 0.008) \; \mu{\rm s}^{-1}$ 
are the corresponding parameters for the diamagnetic signal with $\nu_D = (-0.2585 \pm 0.0009)$ MHz;
the diamagnetic initial phase is held equal to that of muonium.  

All the aerogel targets (even those composed of PMSQ instead of silica) 
yielded the same muonium amplitudes, 
within statistical uncertainties; therefore all such data were combined 
to make a much higher statistics $\mu$SR time spectrum 
from which the important results were extracted.  
For the sum of all data on aerogel targets in the same field, 
the fitted value of $A_{\rm Mu} = 0.155 \pm 0.004$, 
combined with the separate determination of $A_0 = 0.474 \pm 0.004$ 
from an aluminum plate, we can conclude that   
\begin{equation}
 f_{\rm Mu} = \frac{2A_{\rm Mu}}{A_0} = 0.655 \pm 0.018 \; .
\end{equation}
The obtained muonium fraction is larger than the one obtained in the past study ($0.52 \pm 0.01$)~\cite{PTEP2013}. These studies used silica aerogel targets produced in slightly different synthesis conditions. The measurement methods and apparatus were also different. Although the discrepancy of the two muonium fractions is not understood well, it does not affect the muonium emission rates because they were measured independently.

\subsubsection{Muonium polarization in vacuum}

In contrast to the simple exponential for muon decay in all region, 
muon decay in a vacuum region is proportional to the probability 
that the muonium atom is in that region. 
We assumed that the muonium atom is emitted with a velocity 
following the Maxwell-Boltzmann distribution 
and derived the corresponding time distribution 
assuming a travel distance $\ell$.  
The Maxwell-Boltzmann velocity distribution in the $z$ direction is 
\begin{equation}
 {\cal D}(v_z) = \sqrt{\frac{m}{2\pi k_B T}} 
   \exp\left(-\frac{mv_z^2}{2k_B T}\right) = 
   B e^{-b v_z^2} \; ,
\end{equation}
where 
\begin{equation}
B \equiv \sqrt{\frac{m}{2\pi k_B T}} = \sqrt{\frac{b}{\pi}} 
\quad \hbox{\rm and} \quad 
   b \equiv \frac{m}{2k_B T} \; .
\end{equation}
{\sl For a fixed distance\/} $\ell$, that gives a TOF distribution 
\begin{equation}
 {\cal D}(t,\ell) dt = {\cal D}(v_z) dv_z \quad \hbox{\rm or} \quad 
   {\cal D}(t.\ell) = {\cal D}(v_z) \left| \frac{dv_z}{dt} \right| \; ,
\end{equation}
and since $v_z = \ell/t$, $dv_z/dt = -\ell/t^2$, giving 
\begin{equation}
 {\cal D}(t,\ell) = B \ell t^{-2} \exp\left(-C t^{-2}\right) \; ,
\label{eq:RawDist}
\end{equation}
where $C \equiv b \ell^2$. 
Regions 1, 2 and 3 cover $z$ ranges of $10~{\rm mm}-20$ mm, $20~{\rm mm}-30$ mm and $30~{\rm mm}-40$ mm 
from the target, respectively, so that the actual ${\cal D}(t)$ in each region 
should in turn be weighted by the corresponding $\ell$ distribution ${\cal D}(\ell)$: 
$$
 {\cal D}(t) = \int_{\rm region} d\ell \; {\cal D}(\ell) \; {\cal D}(t,\ell) \; .
$$
Especially in region 1, there is also a distribution of emission angles 
to be considered.  We were unable to convert this combination of distributions 
into an algebraic expression for fitting, so we did the next best thing: 
we simply substituted an {\sl average\/} $\langle \ell^2 \rangle$ for $\ell^2$, giving 
\begin{equation}
 C = b \langle \ell^2 \rangle 
\label{eq:C}
\end{equation}
in Eq.~(\ref{eq:RawDist}).

\begin{figure}[htbp] 
	\begin{minipage}{\hsize}
	\begin{center}
\includegraphics[width=0.75\columnwidth,angle=0]{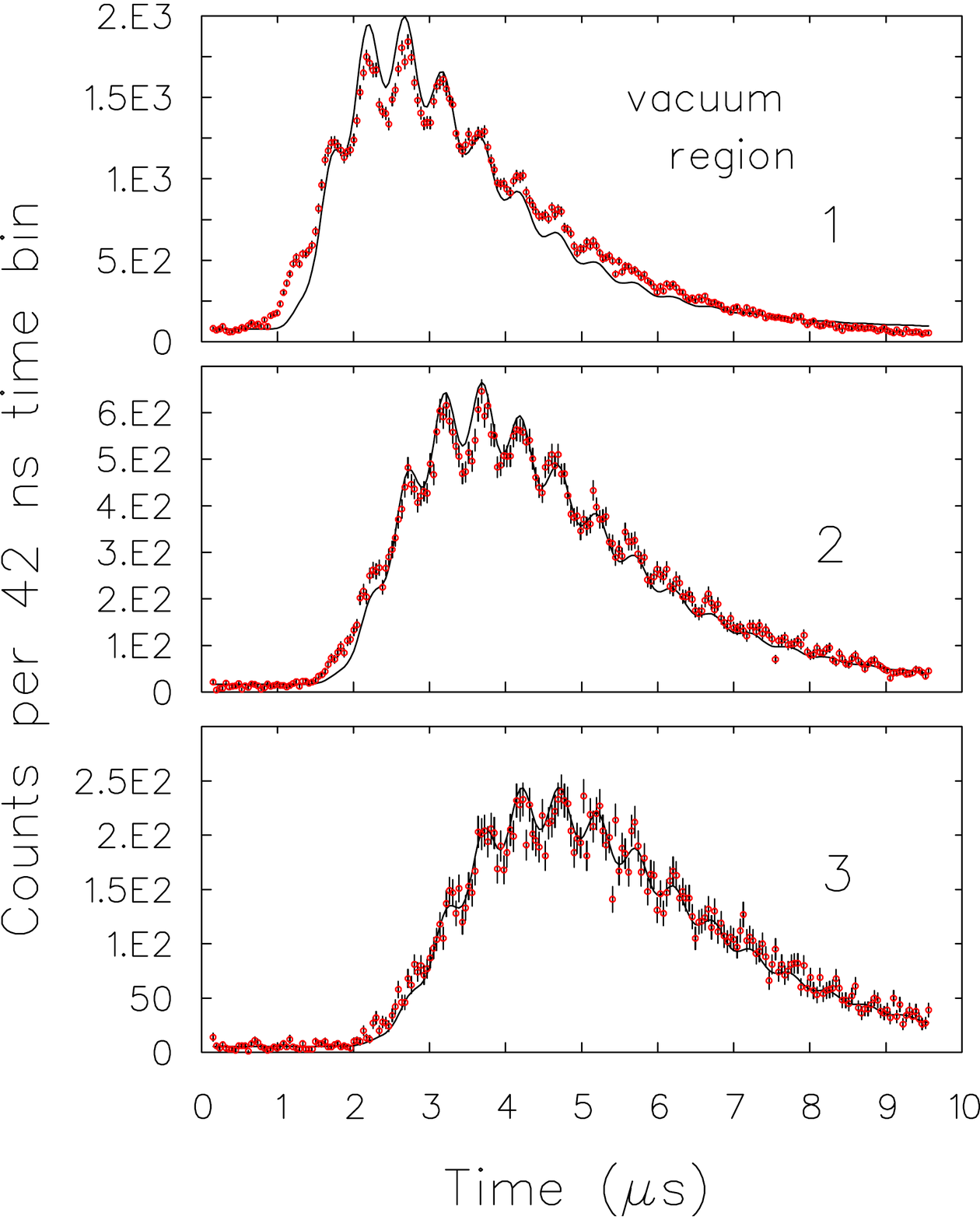}
	\end{center}
	\end{minipage}
\caption{
Fits to region 1 (top), 2 (middle) and 3 (bottom) for the sum of all data
on aerogel targets in ambient field.  
The fit quality is worst in region 1, probably because 
$\ell$ is far from constant across this region, 
while it is much better in region 2 and even better in region 3.  
} 
\label{fig:N2-4} 
\end{figure} 

Fits in regions 1, 2 and 3 to the product of 
Eq.~(\ref{eq:RawDist}) and Eq.~(\ref{eq:NofT}) 
with $BG = 0$ and $A_{\rm D} = 0$ 
(no diamagnetic signal is expected in the vacuum regions) 
are shown in Fig.~\ref{fig:N2-4}.
The time evolution of spectra over the three regions indicates emergence of muonium atoms into vacuum with a thermal velocity distribution.
The time spectra are clearly much as expected though with some mismatch in region 1.
For $T = 300$ K and the mass $m$ of the muonium atom, 
$b = 0.2274\times 10^{-7}$~s$^2$/m$^2$.  
These fits yield $C$ values that can be combined with $b$ 
to obtain $\ell = \sqrt{C/b}$, as shown in Table~\ref{tbl:ell}.  

\begin{table}[htb]
\caption{
 The fitted velocity spectrum parameter $C$ and the rms travel distance 
 $\ell_{\rm rms} = \sqrt{C/b}$ obtained for each vacuum region of the listed $z$ range 
 from the fits described in the text. 
}
\vspace*{4mm} 
\centering
\small
\begin{tabular}{ccrc}
\hline\hline
Region & $z$ [mm] & $C$ [$\mu$s$^2$] & $\ell_{\rm rms}$ [mm] \\
\hline\hline
 1 & $10-20$ & $9.54 \pm 0.04$ & $20.48 \pm 0.05$ \\
 2 & $20-30$ & $22.69 \pm 0.13$ & $31.59 \pm 0.09$ \\  
 3 & $30-40$ & $40.5\ \, \pm 0.3\ \, $ & $42.19 \pm 0.16$ \\
\hline\hline
\end{tabular}
\label{tbl:ell}
\end{table}

We see clear muonium precession signal in the spectra. In principle, we can use the precession amplitude to extract the muonium polarization in vacuum. However, we found it was difficult to extract a reliable value due to the following reasons. First, in order to show spectra with best statistics, the spectra were for the sum of measurements under ambient field of about 0.14 mT and the field axis had large tilt to the $y$-axis. This causes a tilt of the spin precession plane and thus the detector sensitivity to the polarization could be much reduced compared to that in Fig. 11 under the applied field of 1.92 mT.
Second, a field variation due to the disturbance of experimental environment found
over the measurements (in space and in time) causes significant relaxation of the precession and it is difficult to extrapolate the result back to the amplitude at time zero.
Third, the shape of the TOF spectrum calculated from the Maxwell-Boltzmann distribution is only an approximation, since muonium atoms emerge at different angles; this interferes with the estimation of the initial muonium polarization.  So, at the present stage of the analysis, we cannot conclude any reliable value for the muonium polarization.
However, no depolarization mechanism is expected in the aerogel target nor in vacuum and we observed the muonium precession in vacuum as expected. A quantitative confirmation of the muonium polarization in vacuum remains to be studied.

\section{Measurements and projected rates at J-PARC}
\subsection{Beam condition and experimental setup}

The muonium emission rates in vacuum were also measured at J-PARC for five aerogel targets that include a flat and three ablated ones used in the TRIUMF measurements.
The emission rates measured at TRIUMF for the four targets can be projected to the corresponding rates at J-PARC, taking into account the different experimental conditions.
A reasonable projection, if available, can demonstrate the procedure to apply the knowledge on the muonium emission obtained under the experimental condition at TRIUMF to the simulation for the J-PARC muon $g$-2/EDM experiment.

The measurements were carried out with a pulsed subsurface muon beam at the D2 area of Muon Science Establishment (MUSE)~\cite{MUSE} in J-PARC MLF.
The MUSE beamlines deliver intense pulsed muons at the repetition rate of 25 Hz, having two bunches 600 ns apart with the bunch width of 50 ns (RMS) for each.
The beam momentum at D-line has a spread of about 4\% (RMS) based on a simulation and was tuned to be 23.7 MeV/$c$ as done in the TRIUMF measurements for the half-stop condition.

The experimental setup was similar to that in the TRIUMF measurement shown in Fig.~\ref{fig:setup_chamber}, but the pieces of apparatus are different except the target chamber and holder.
No coils were used since only the muonium emission rates were measured.
The pulsed beam timing distributed from the accelerator, instead of the beam counter, was used to start the data taking cycle and an additional lead collimator with the hole diameter of 10~mm was used, instead of the veto counter. A degrader of layered polyimide films was set at the exit of the collimator in order to have a similar amount of substances to the TRIUMF measurements compensating the removal of the beam counter.
The positron detection system consisted of two layers of scintillation-fiber hodoscope (S1, S2), followed by a polyethylene absorber, and a 4$\times$4 matrix of 20~mm-thick scintillators (S3). A positron track was determined from the hit positions of S1, S2 and S3, and was extrapolated to the $y-z$ plane to infer the decay position in the same manner as in the TRIUMF measurements.

\subsection{Muonium emission rates}

The muonium emission rates were evaluated in a similar manner as in the TRIUMF measurements.
The resolution of the track extrapolation was evaluated to be about 9.0 mm (FWHM) from the silica plate data, giving a wider distribution of the target decay background over the $z$-range compared to that in the TRIUMF measurements.
To ensure the signal dominance in the emission rate evaluation, a smaller vacuum region of 20 mm $< z <$ 40 mm was adopted in the J-PARC measurements, while the fiducial ``all'' region was kept to be the same ($-20$ mm $< z <$ 40 mm).
The contributions of the target decays and outlier events in vacuum were modeled using the silica plate data. The fraction of those backgrounds in the vacuum region was considered to be the same for the silica plate and aerogel targets at the half-stop condition.
Therefore, the observed event rate of the silica plate, $(4.80 \pm 0.31)$\textperthousand, was subtracted from those of the other targets.

\begin{table}[bp]
\caption{
Summary of the muonium emission rates in vacuum ($R_{\rm Mu}^{\rm vac}$) and their statistical uncertainties measured at J-PARC D-line as described in the text.
}
\label{tab:yields_jparc}
\centering
\small
\begin{tabular}{ccr}
\hline\hline
Ablation & Target & \multicolumn{1}{c}{$R_{\rm Mu}^{\rm vac}$ [\textperthousand]}\rule[-1mm]{0mm}{4mm}\\
\hline\hline
Flat  & S08 & $1.20 \pm 0.70$ \\
\cdashline{1-1}
\multirow{4}{*}{Holes} & S01 & $6.14 \pm 0.70$ \\
 & S03 & $5.82 \pm 0.65$ \\
 & S04 & $5.09 \pm 0.49$ \\
 & S05 & $4.10 \pm 0.50$ \\ 
\hline\hline
\end{tabular}
\end{table}

The obtained muonium emission rates after the background subtraction are listed in Table~\ref{tab:yields_jparc} together with the statistical uncertainties. The dominant systematic uncertainty comes from the target position variation caused by a mechanical clearance of 0.6~mm in the target holder mounting used in the J-PARC measurements. Remounting of the target holder at the target replacement can give a possible variation of the target position within the clearance. The observed event rates were re-evaluated varying the vacuum region by $\pm 0.3$ mm, and the resulting relative changes from the nominal rates, $-10$\% to $+5$\%, were assigned as the systematic uncertainty.

\begin{figure}[htbp] 
\begin{center} 
\includegraphics[width=0.7\hsize, angle=0]{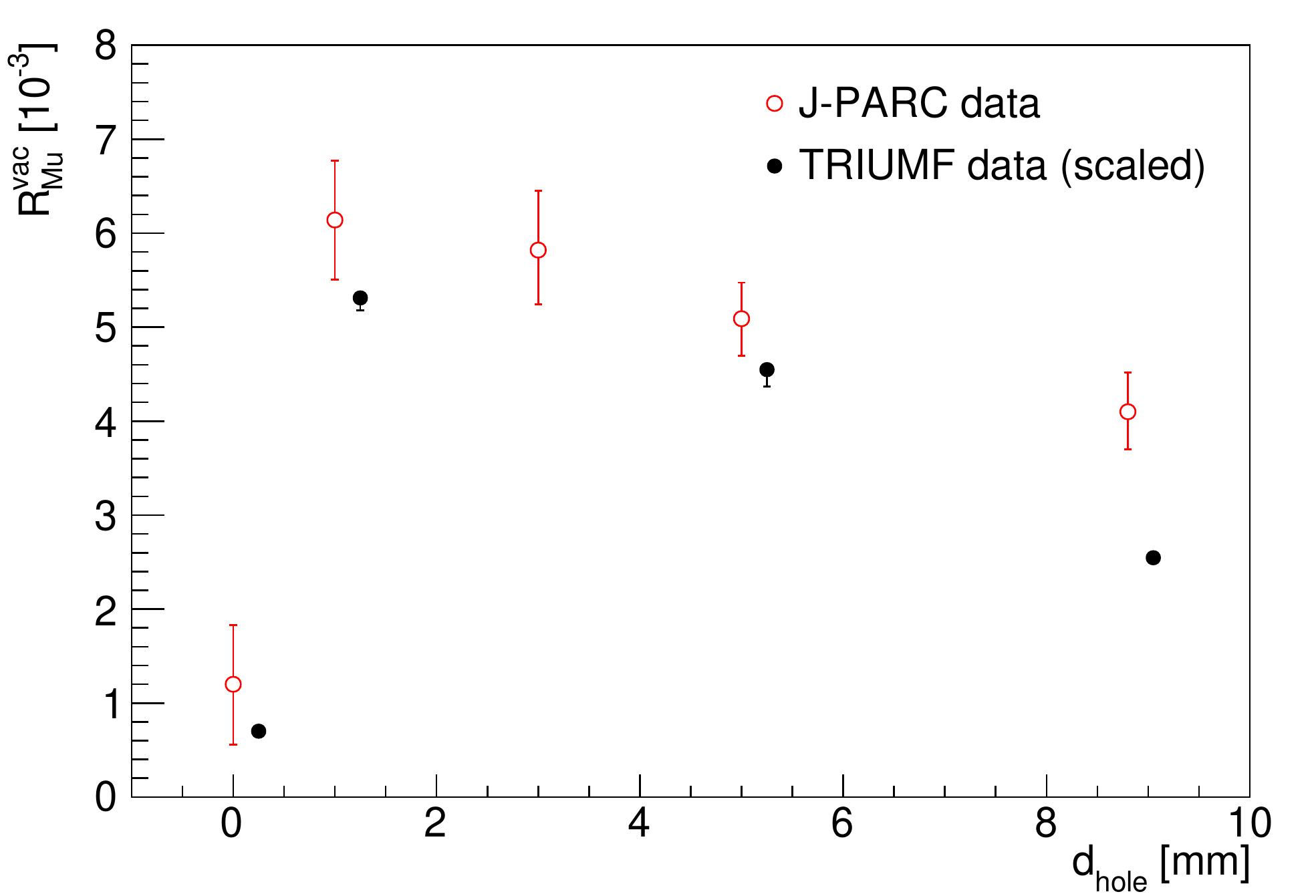}
\caption{
The muonium emission rates in vacuum ($R_{\rm Mu}^{\rm vac}$) measured at J-PARC D-line as a function of hole depth ($d_{\rm hole}$) (red open circle).
The error bars represent the statistical and systematic uncertainties combined in quadrature, excluding the statistical uncertainty of the silica plate data which was given in the background subtraction and common for all the rates.
The corresponding TRIUMF measurements available for the four targets are projected to the experimental condition at J-PARC as described in the text and are also shown (black closed circle) with a horizontal offset for presentation convenience. Here, the error bars represent the combined uncertainties in the same manner with the J-PARC data; the common projection uncertainty of about $\pm 10$\% are also not included.
} 
\label{fig:yield_dep}
\end{center} 
\end{figure} 

The four aerogel targets with laser ablation have identical hole structure parameters except for the depth. Their emission rates are compared in Fig.~\ref{fig:yield_dep} as a function of the hole depth. A trend of the rate reduction as the holes become deeper can be seen as in the TRIUMF measurements.
Here, an error bar represents the quadratic sum of the statistical and systematic uncertainties excluding the statistical uncertainty of the silica plate data which was given in the background subtraction and common for all the rates.

\subsection{Projection of the TRIUMF measurements}

The corresponding TRIUMF measurements for the flat and three ablated targets are also projected in Fig.~\ref{fig:yield_dep}, taking into account the major differences between the two sets of the measurements.
One of the major differences exists in the beam momentum spread.
The muonium diffusion model~\cite{PTEP2013} indicates that the emission rate is proportional to the muon stopping density and, hence, is inversely proportional to the range spread in the muon stopping, which is largely correlated with the beam momentum spread.
To account for the different beam momentum spreads, therefore, the rates in the TRIUMF measurements need to be scaled in the projection by their ratio.
The range spread influences the slope of the positron yield variation of the target decays over the beam momentum, shown in Fig.~\ref{fig:momscan_noabl}.
Comparing the variation slopes observed in the TRIUMF and J-PARC measurements, the momentum spread ratio was evaluated to be 0.6.

The other major difference exists in the resolution of the track extrapolation. The difference leads to the smaller vacuum region in the J-PARC measurements as described earlier. The ratio of the positron yields in the two regional definitions can be taken as the scale factor to account for the vacuum region difference. The scale factor was evaluated to be 0.36 using the TRIUMF data.
The resolution difference also results in the event migration over a given region.
This effect was calculated based on the observed track extrapolation resolutions using the TRIUMF data, giving the scale factor of 1.1.

The projected rates of the TRIUMF measurements to the J-PARC condition were obtained applying the three scale factors at the combined accuracy of $\pm 10$\% level.
The similarity of their rate reduction trends to the hole depth can be seen clearly in Fig.~\ref{fig:yield_dep}. The projected TRIUMF rates and the measured J-PARC rates,
except the ones with the through-hole ablation ($d_{\rm hole} = 8.8$ mm), agree at $1\sigma$ level considering the projection accuracy ($\pm$10\%) and the uncertainty of the J-PARC measurements ($\pm$10\%).
While more details can be considered in the projection, the treatment of the different beam condition seems to be reasonable.

\subsection{Expected thermal muonium rate in the J-PARC muon $g$-2/EDM experiment}

The projection demonstrates that the beam property dominates the initial condition of the muonium diffusion and emission processes, while the processes themselves can be the same between the experimental conditions at TRIUMF and J-PARC; the other scalings account for the difference on how to measure the emission rates.
This demonstration supports the application of the model of the two processes developed in the study of past measurements at TRIUMF~\cite{PTEP2013} in the simulation to estimate the expected rate of thermal muonium atoms available for laser ionization in the J-PARC muon $g$-2/EDM experiment.

The simulation includes the surface muon transport through the beamline, the muon stopping and muonium diffusion in the aerogel target, and the muonium emission from the target followed by the travel in vacuum.
The simulation incorporates the design of MUSE H-line~\cite{Hline} under construction, at which the J-PARC muon $g$-2/EDM experiment will be launched.
The momentum spread of the surface muons is expected to be 5\% (RMS) there.
The expected surface muon rate at H-line was estimated from the the measured rate at D-line, which shares the muon production target at the symmetric configuration with regard to the primary proton beam.
The beam transport efficiencies at H-line and D-line were estimated using the G4beamline program~\cite{G4beamline}, and their ratio was used in the rate estimation.
The muon stopping was simulated with the Geant4 toolkit. All the stopped muons were assumed to form muonium atoms in the simulation and the formation fraction was applied in the rate estimation.
The muonium diffusion inside the target was modeled with the Maxwell-Boltzmann velocity distribution at room temperature.
The angular distribution of muonium emission was modeled as proportional to $\cos\theta$, where $\theta$ is the polar angle of the muonium emission direction with respect to the emission face normal.
The laser ionization region was defined to be $\pm 100$~mm and $\pm 25$~mm regarding the beam axis in the horizontal and vertical directions, respectively, and 5~mm along the beam axis from the target emission face. 
The simulated time-evolution of the muonium travel shows the highest population density in the laser ionization region at the muonium arrival time of around 1~$\mu$s from the averaged arrival time of the surface muons over the two bunch structure.
The expected rate of the muonium atoms for laser ionization is estimated to be $3.4\times 10^{-3}$ per incident surface muon~\cite{E34}.

\section{Summary}

The muonium emission from the muonium production targets with various laser-ablated structures on their surfaces was studied at room temperature as the development of the thermal muon source for the muon $g$-2/EDM experiment at J-PARC.
The study was intended to understand better the underlying processes leading to muonium emission with laser ablation, to find the optimal ablation structure, and to confirm the polarization of muonium following emission into vacuum.
The muonium emission rates in vacuum and precession amplitudes were measured irradiating the subsurface muon beams at TRIUMF and J-PARC on silica aerogel targets with a typical density of 23.6 mg/cm$^3$ and PMSQ-gel targets with higher densities.
Laser ablation was applied to produce holes or grooves with a diameter or width, respectively, of $0.07\ {\rm mm} - 0.5$ mm varying the fractions of ablation opening and a depth of $0.3\ {\rm mm} - 5$ mm, except for some extreme conditions.
The measured emission rate tends to be higher for larger fractions of ablation opening, rather than the larger absolute surface area of ablation, and for shallower depths with a possible optimum at less than 2 mm.
More than a few ablation structures reached the emission rates similar to the highest achieved in the past measurements, which would be applicable to an early phase of the J-PARC muon $g$-2/EDM experiment.
The emission rate was found to be stable at least for a couple of days with no physical damage for the long-time exposure to the high-vacuum environment.
The measurements of spin precession amplitudes for the produced muonium atoms and remaining muons in a magnetic field determined a muonium formation fraction of $(65.5 \pm 1.8)$\%.
The precession of the polarized muonium atoms was also observed clearly in vacuum.
A projection of the emission rates measured at TRIUMF to the corresponding rates at J-PARC was demonstrated taking the different beam condition into account reasonably. The projected rates agree with the ones measured at J-PARC at $1\sigma$ level, except the ones with the through-hole ablation.
This projection procedure was applied to the estimation of the expected rate of thermal muonium atoms available for laser ionization in the muon $g$-2/EDM experiment at J-PARC.

\section*{Acknowledgement}
The authors are pleased to acknowledge the support from TRIUMF to provide a stable beam during the experiment. Special thanks go to R. Henderson, R. Openshaw, G. Sheffer, and M. Goyette from the TRIUMF Detector Facility. We also thank D. Arseneau, G. Morris, B. Hitti, R. Abasalti, and D. Vyas of the TRIUMF Materials and Molecular Science Facility. Continuous support given by P. Strasser from KEK has been a great help in the experiment at J-PARC MLF. We are grateful for useful discussion with E. Torikai and her support.
We are thankful to H. Kawai from Chiba University for his support to produce silica aeogel samples. We are also thankful to M. Aizawa from tiem factory Inc. for the production of PMSQ gels. We would like to thank C. Herrmann for acquiring the He-ion microscope image at the 4D LABS facility at Simon Fraser University.
The research was supported in part by MEXT KAKENHI Grant Numbers 23108005 and 23108001, JSPS KAKENHI Grant Numbers JP18H05226 and JP17K05466 (Japan), NSERC Discovery Grant (Canada) and Canadian Foundation for Innovation (CFI).



\begin{thebibliography}{9}

\bibitem{muSR}
J.~H.~Brewer, ``Muon Spin Rotation/Relaxation/Resonance'', Encyclopedia of Applied Physics {\bf 11}, 23-53, VCH Publishers, Inc., (1994);
%
A.~Yaouanc and P.~D.~de R{\'e}otier, ``Muon Spin Rotation, Relaxation, and Resonance: Applications to Condensed Matter'',
International Series of Monographs on Physics {\bf 147}, 1-504, Oxford University Press (2010);
%
J.~H.~Brewer, ``Methods and Applications of $\mu$SR'', Modern Muon Physics: Selected Issues, Nova Science Publishers, Inc. (2020).

\bibitem{SurfaceChem}
R.~F.~Marzke {\it et al.},
``Magnetic Susceptibility, Proton NMR and Muon Spin Rotation ($\mu$SR) Studies of an Unsupported Platinum Catalyst with Adsorbed H and O'', {\bf 1}, 647, Plenum Press (1983);
%
R.~F.~Marzke {\it et al.},
Chem. Phys. Lett. {\bf 120}, 6 (1985);
%
R.~F.~Marzke {\it et al.},
Ultramicroscopy {\bf 20}, 161 (1986);
%
J.~R.~Kempton {\it et al.}, 
Hyperfine Int. {\bf 65}, 811 (1990);
%
M.~Schwager {\it et al.},
Hyperfine Int. {\bf 87}, 859 (1994);
%
M.~H.~Dehn {\it et al.},
J. Chem. Phys. {\bf 145}, 181102 (2016);
%
M.~H.~Dehn {\it et al.},
Appl. Phys. Lett. {\bf 112}, 053105 (2018);
%
D.~G.~Fleming {\it et al.}, 
J. Phys. Chem. C {\bf 123}, 27628 (2019).

\bibitem{MuLambShift}
C.~J.~Oram {\it et al.}, Phys. Rev. Lett. {\bf 52}, 910 (1984);
A.~Badertscher {\it et al.}, Phys. Rev. Lett. {\bf 52}, 914 (1984);
K.~A.~Woodle {\it et al.}, Phys. Rev. {\bf 41}, 94 (1990).

\bibitem{MuHFS}
W.~Liu {\it et al.}, Phys. Rev. Lett. {\bf 82}, 711 (1999);
P.~Strasser {\it et al.}, Hyperfine Interact. {\bf 237}, 124 (2016).

\bibitem{Mu1S2S}
S.~Chu {\it et al.}, Phys. Rev. Lett. {\bf 60}, 101 (1988);
F.~E.~Maas {\it et al.}, Phys. Lett. A {\bf 187}, 247 (1994);
V.~Meyer {\it et al.}, Phys. Rev. Lett {\bf 84}, 1136 (2000);
I.~Fan {\it et al.}, Phys. Rev. A {\bf 89}, 032513 (2014);
P.~Crivelli, Hyperfine Interact. {\bf 239}, 49 (2018).

\bibitem{LEM}
T.~Prokscha {\it et al.},
Appl. Surf. Sci. {\bf 172}, 235 (2001).

\bibitem{MuCool}
D.~Taqqu, Phys. Rev. Lett. {\bf 97}, 194801 (2006);
Y.~Bao {\it et al.},
Phys. Rev. Lett. {\bf 112}, 224801 (2014).

\bibitem{MICE}
D.~Adams {\it et al.}, Eur. Phys. J. C 2019, {\bf 79}, 257 (2019).

\bibitem{LaserRes}
K.~Nagamine et al.,
Phys. Rev. Lett. {\bf 74}, 4811 (1995).

\bibitem{LaserResRAL}
P.~Bakule {\it et al.}, Nucl. Instrum. Methods Phys. Res., Sect. B {\bf 266}, 335 (2008).

\bibitem{LymanAlpha}
N.~Saito {\it et al.}, Opt. Express {\bf 24}, 7566 (2016).

\bibitem{Uline}
A.~Pant {\it et al.}, JPS Conf. Proc. {\bf 21}, 011060 (2018).

\bibitem{E34}
M.~Abe {\it et al.}, Prog. Theor. Exp. Phys. {\bf 2019}, 053C02.

\bibitem{PTEP2013}
P.~Bakule {\it et al.}, Prog. Theor. Exp. Phys. {\bf 2013}, 103C01.

\bibitem{PTEP2014}
G.~A.~Beer {\it et al.}, Prog. Theor. Exp. Phys. {\bf 2014}, 091C01.

\bibitem{SilicaAerogel}
M.~Tabata {\it et al.}, Nucl. Instrum. Methods Phys. Res., Sect. A {\bf 668}, 64 (2012).

\bibitem{PMSQ}
K.~Kanamori {\it et al.}, Adv. Mater. {\bf 2007}, 19, 1589.

\bibitem{Coherent} Coherent Legend. The repetition rate (1 kHz default) and energy (2 mJ/pulse max.) were varied for the ablation work.

\bibitem{Zeiss} Zeiss LSM 510.
Information of the latest-generation model can be found in the following URL:
https://www.zeiss.com/microscopy/int/products/confocal-microscopes.html.

\bibitem{Subsurface}
A.~Badertscher {\it et al.}, Nucl. Instrum. Methods Phys. Res. Sect. A {\bf 238}, 200 (1985).

\bibitem{MWDC}
R.~S.~Henderson {\it et al.}, Nucl. Instrum. Methods Phys. Res., Sect. A {\bf 548}, 306 (2005).

\bibitem{Geant4}
S.~Agostinelli {\it et al.}, Nucl. Instrum. Meth. A {\bf 506}, 250 (2003);
J.~Allison {\it et al.}, IEEE Trans. Nucl. Sci. {\bf 53}, 270 (2006);
J.~Allison {\it et al.}, Nucl. Instrum. Meth. A {\bf 835}, 186 (2016).

\bibitem{Thermalization} 
J.~H.~Brewer, K.~M.~Crowe, F.~N.~Gygax and A.~Schenck in ``Muon Physics'', Academic Press, Vol. III;
K.~Nagamine, ``Introductory Muon Science'', Cambridge University Press (2007).

\bibitem{DMF}
D.~G.~Eshchenko {\it et al.},
Phys. Rev. B {\bf 66}, 35105 (2002).

\bibitem{Mariam1982}
F.~G.~Mariam {\it et al.},
Phys. Rev. Lett. {\bf 49}, 993 (1982).   

\bibitem{MUSE}
W.~Higemoto {\it et al.},
Quantum Beam Sci. {\bf 1}, 11 (2017).

\bibitem{Hline}
N.~Kawamura {\it et al.},
Prog. Theor. Exp. Phys. 2018, 113G01 (2018).




\bibitem{G4beamline}
Information is avaialble in the following URL: http://g4beamline.muonsinc.com.

\end{thebibliography}
%


\end{document}